\documentclass[twocolumn,aps,showpacs,pre]{revtex4}
\usepackage{amsmath}
\usepackage{graphicx}
\usepackage{float}
\usepackage{subfigure}
\usepackage{xcolor}

\begin{document}

\title{Non-equilibrium view of the amorphous solidification \\ of  liquids with competing interactions}

\author{Ana Gabriela Carretas-Talamante$^{1}$, Jes\'us Benigno Zepeda-L\'opez$^{1}$, Edilio L\'azaro-L\'azaro$^{1}$,  \\  Luis Fernando Elizondo-Aguilera$^{2}$, and  Magdaleno Medina-Noyola$^{1,(a)}$} 

\affiliation{$^{1}$ Instituto de F\'{\i}sica {\sl ``Manuel Sandoval Vallarta"}, Universidad Aut\'{o}noma de San Luis Potos\'{\i}, \'{A}lvaro Obreg\'{o}n 64, 78000 San Luis Potos\'{\i}, SLP, M\'{e}xico}

\affiliation{$^2$ Instituto de F\'isica, Benem\'erita Universidad 
Aut\'onoma de Puebla, Apartado Postal J-48 72570, Puebla, M\'exico}
\email{(a) medina@ifisica.uaslp.mx}

\date{\today}

\begin{abstract}
The interplay between short-range attractions and long-range repulsions (SALR) characterizes the so-called liquids with competing interactions, which are known to exhibit a variety of equilibrium and non-equilibrium phases. The theoretical description of the phenomenology associated with glassy or gel states in these systems has to take into account both the presence of thermodynamic instabilities (such as those defining the spinodal line and the so called $\lambda$ line) and the limited capability to describe genuine non- equilibrium processes from first principles. Here, we report the first application of the non-equilibrium self-consistent generalized Langevin equation theory to the description of the dynamical arrest processes that occur in SALR systems after being instantaneously quenched into a state point in the regions of thermodynamic instability. The physical scenario predicted by this theory reveals an amazing interplay between the thermodynamically driven instabilities, favoring equilibrium macro- and micro-phase separation, and the kinetic arrest mechanisms, favoring non-equilibrium amorphous solidification of the liquid into an unexpected variety of glass and gel states.
\end{abstract}


\maketitle

\section{Introduction.}\label{section1}
This work reports the first systematic application of the non-equilibrium self-consistent generalized Langevin equation theory \cite{nescgle1, nescgle2, nescgle3}, to the description of non-equilibrium  arrested states in fluids with competing short-range attractions and long-range repulsions (SALR) \cite{LiuXi2019,liuJPCB}. Such arrested states may prevent the fluid from reaching the experimentally elusive \cite{zhuangcharbonneauJCP16} ordered phases expected at  thermodynamic equilibrium conditions in the low-density  low-temperature regime of these systems \cite{ruizzaccarelli}. 

As a context, let us first recall that van der Waals (vdW) molecular explanation of the gas-liquid coexistence had in mind a model fluid of spherical particles interacting by an excluded-volume repulsion plus a weaker short-ranged attraction \cite{vdw,widom}. 
In choosing this simple model of a fluid, van der Waals was fortunate enough since, for example, innocently adding a soft longer-ranged repulsive interaction immediately leads to a different class of systems, whose equilibrium phase behavior turns out to be far more complex. This more general class, characterized by the excluded-volume potential plus the competing SALR interactions, bears an enormous relevance in colloidal and soft materials. Familiar physical realizations of SALR conditions may be represented, for example, by the effective interaction between two charged particles (colloids, proteins, or macroions) in aqueous solution \cite{verweyoverbeek,derjaguinlandau} or between two colloids in colloid-polymer mixtures \cite{asakura}. 

Different approaches have been applied to determine the equilibrium phase diagram of SALR systems, including theoretical methods (integral equations  
\cite{seargelbart,pinijealinparolareatto,imperioreatto04,LiuChenChenJCP05}, density functional theory \cite{archerpinievansreatto07,archerionescu3Djpcm08,archer2Dpre08,chackochalmersarcher2Djcp15}, and field-theoretical models described by Ginzburg-Landau Hamiltonians \cite{SchmalianWolynesPRL00,TarziaConiglioPRL06,TarziaConiglioPRE07}) and Monte Carlo computer simulations \cite{candiaconiglio,imperioreato,archerwilding,zhuangzhangcharbonneau,godfrin}. As recently and thoroughly reviewed  \cite{LiuXi2019,ruizzaccarelli}, the picture that emerges predicts a rich and diverse phenomenology, which includes the appearance at low densities of an equilibrium fluid phase of finite-sized clusters, the coexistence between disordered equilibrium  (gas and liquid) phases and ordered (``modulated'') inhomogeneous phases, and the possibility that equilibrium microphase separation preempts the usual vdW gas-liquid coexistence. 

These different phenomenological features correspond in general to different combinations of the short-range attraction and the long-range repulsion. A major challenge is, of course, to establish which salient features correspond to which regime in the four-dimensional parameter space spanned by the  intensity and range of the two competing interactions. After careful analysis of experimental, simulation and theoretical results, Liu and Xi \cite{LiuXi2019} suggested that most of SALR systems can be grouped in three different regimes, identified by the ratio between ranges of the attractive and repulsive interactions. They noticed that most model experimental SALR systems studied, belong to a similar type, referred to as  type I SALR systems, whereas  many simulation works are type II SALR systems. As discussed in detail in  \cite{LiuXi2019}, the difference between different types of SALR systems has strong impact on the final equilibrium phase  diagram. As specified below, however, the present work  will focus on still a third SALR regime, referred to in \cite{LiuXi2019} as type III SALR systems.



Given this equilibrium scenario, the next most important issue is its actual experimental observability. This concern arises from the possible emergence of kinetic barriers to equilibration, which may lead to conditions of dynamic arrest,  as suggested by early theoretical considerations \cite{groenewoldkegel2001,groenewoldkegel2004,wuliuchencao,LiuChenChenJCP05} and by molecular dynamics (MD) studies \cite{conigliodearcangelis,sciortinomossa,mossasciortino,sciortinotartaglia,charbonneaureichmanPRE07,BollingerTruskettSoftMatter2017}. The latter have confirmed, for example, the presence of a phase of stable, freely diffusing clusters of particles and of non-equilibrium phases of disordered arrested states. Beyond its fundamental relevance, this knowledge is important for many practical purposes, such as designing rules for the assembly of porous mesophases \cite{LindquistTruskettSoftMatter2017}. In contrast with Montecarlo simulations, molecular dynamics (and Brownian dynamics (BD)) simulations mimic realistic trajectories in phase space, whose ensemble averages directly relate with the real dynamical phenomenology, actually observed since early and careful  experimental studies \cite{sedgwick,stradner,bordi,campbell,DibbleKoganSolomonPRE06,KlixRoyalTanakaPRL2010}.

Regarding the theoretical efforts to understand the formation of amorphous arrested states in SALR systems, let us mention the analysis aimed at determining if a glass transition exists in field-theoretical models. These efforts include the use of the replica approach \cite{mezardparisiPRL99,SchmalianWolynesPRL00,TarziaConiglioPRL06,TarziaConiglioPRE07} and of the Langevin-equation version of these models \cite{GroussonKrakoviackPRE02,GeisslerReichmanPRE04}. Early applications of mode coupling theory (MCT) \cite{goetze1,goetze2} to SALR systems were also reported by Grousson et al. \cite{GroussonKrakoviackPRE02}  and by  Geissler and Reichman \cite{GeisslerReichmanPRE04}, although only for  field-theoretical models. 

However, developing a general fundamental microscopic theory of dynamical arrest in structural glasses, which explains the glass transition and extends the van der Waals picture to non-equilibrium conditions and to more complex interactions (such as SALR systems), has been the purpose of relevant work over the last half a century \cite{berthierbiroli}. In this context, one should highlight the first-principles description of the dynamic properties of fluids near conditions of dynamic arrest provided by MCT  \cite{goetze1,goetze2}, which predicts the location of the transition from equilibrium-fluid to dynamically-arrested states. As early as in 2004, Wu et al. \cite{wuliuchencao} discussed the application of MCT to the hard-sphere plus double Yukawa SALR type II interaction, predicting its fluid-to-glass transition diagram in the high-density high-temperature regime. Unfortunately, this work did not explore in detail the opposite (low-density low-temperature) regime.


Starting this exploration is precisely the purpose of the present work, motivated by the need to understand if the emergence of kinetic arrest and non-ergodicity might be the source of the experimental elusiveness of the predicted ordered (or ``modulated'') equilibrium phases, which characterizes one specific SALR regime. In contrast with the work of Wu et al., however, our work will not be based on MCT, but on the more recent theory referred to as the self-consistent generalized Langevin equation (SCGLE) theory of colloid dynamics \cite{scgle1,scgle2,scgle3,scgle4,scgle5} and dynamical arrest \cite{arrest1,arrest2,arrest3}, which in most aspects is analogous to MCT  \cite{thvg_elizondo}. 

In reality, however, the SCGLE theory, just like MCT, bears a fundamental constraint to equilibrium conditions, thus impeding the description of essential fingerprints of  dynamic arrest, such as the aging of glass- and gel-forming liquids. The route of escape from this limitation, however, was provided in Ref. \cite{nescgle0}, which proposed a far-from-equilibrium extension of the Onsager theory of irreversible processes  \cite{onsager1,onsager2} and the Onsager-Machlup theory of thermal fluctuations \cite{onsagermachlup1,onsagermachlup2}, leading to the general theory of irreversible processes in liquids, referred to as  the non-equilibrium self-consistent generalized Langevin equation (NE-SCGLE) theory \cite{nescgle1}. This more general approach contains as a particular case the original equilibrium SCGLE theory, which is thus enriched by the non-equilibrium kinetic perspective required to describe non-stationary processes \cite{peredo1}. 

Therefore, the present study will be based on  
the NE-SCGLE, a theory that has solidly demonstrated its ability to predict some of the most relevant universal signatures of both, the glass and gel transitions, including aging effects, as well as very specific features reflecting the particular role of the molecular interactions involved in the explicit systems considered so far. For instance, for systems involving only excluded volume interactions, this theory accurately describes the process of formation of high-density hard-sphere-\emph{like} glasses \cite{nescgle2,nescgle3,gabriel,nescgle5}, whereas for liquids with excluded volume plus attractive interactions (i.e., those in vdW's mind) it predicts the formation of sponge-like gels and porous glasses by arrested spinodal decomposition \cite{olais1,olais2,nescgle8,zepeda}. Extended to  multi-component systems \cite{nescgle4,nescgle5} the NE-SCGLE theory opens the possibility of describing the aging of  ``double'' and ``single'' glasses in mixtures of neutral \cite{voigtmanndoubleglasses,arrest2,lazaro} and charged \cite{arrest3,portadajcppedro} particles; the initial steps in this direction are highly encouraging \cite{lazaro}. Similarly, its extension to liquids formed by particles interacting by non-radially symmetric forces \cite{gory1,gory2,peredo2}, accurately predicts the non-equilibrium coupled translational and rotational dynamic arrest observed in simulations \cite{gory3}.

In this work we start a systematic application of the theoretical infrastructure of the NE-SCGLE theory to the same model of SALR colloidal fluid studied by Wu et al. \cite{wuliuchencao}, namely, a three-dimensional fluid of hard spheres (HS) of diameter $\sigma$, interacting through a total pair potential $u(r)= u^{HS} (r) + u^{DY}(r)$, where $u^{HS} (r)$ is the hard-sphere potential and $u^{DY}(r)$ is the sum of two competing Yukawa interactions, 
\begin{equation}
u^{DY}(r) = -\epsilon_1 \frac{\exp[-z_1(r/\sigma-1)]}{r/\sigma}+ \epsilon_2 \frac{\exp[-z_2(r/\sigma-1)]}{r/\sigma}.
\label{hsdy}
\end{equation}
This model system will be referred to as the ``hard-sphere plus double Yukawa'' (HSDY) fluid, whose equilibrium phase diagram was outlined by Archer and collaborators \cite{archerpinievansreatto07,archerionescu3Djpcm08} using density functional theory within a random phase approximation for the free energy, an approximation also employed in this work. 

In its simplest version, the NE-SCGLE theory is summarized by a set of 
equations describing the irreversible evolution of the non-equilibrium structural and dynamical properties of an instantaneously quenched glass-forming liquid. These properties include the non-equilibrium structure factor (NESF) $S(k;t)$, and the \emph{collective} and \emph{self} intermediate scattering functions (NEISF) $F(k,\tau; t)$ and (self-NEISF) $F_S(k,\tau; t)$, from which the diffusion coefficient, relaxation times, and rheological properties can be derived. Each of these properties depend on the final density and temperature, and on the waiting time $t$, after the quench, as well as on the various parameters characterizing the system (such as the interaction parameters $\epsilon_1,\ \epsilon_2,\ z_1$, and $z_2$). Thus, the analysis of this multidimensional dependence, in its various regimes, will require more than one detailed report, and is completely out of the scope of the present work. Hence, here we shall only summarize the conceptual and practical infrastructure needed for such analysis and will illustrate its use focusing mostly on the structural properties, represented by the NESF $S(k;t)$ and by the corresponding non-equilibrium radial distribution function (NERDF) $g(r;t)$.

For clarity, in Section \ref{section2} we summarize the relevant aspects of the NE-SCGLE theory and briefly explain the procedure for its application to this SALR model fluid. This includes introducing the mean field approximation of the free energy density functional, which determines the main features of the \emph{equilibrium} structure and phase behavior of the system, and is also the fundamental thermodynamic input of the NE-SCGLE equations. Starting with Section  \ref{section3}, we restrict ourselves to the particular regime of long-ranged repulsions ($z_1\geq1> z_2$). This election of parameters will limit this work to exploring only type III SALR system \cite{LiuXi2019}, for which we will discuss what we refer to as the ``glass transition diagram''; this is the NE-SCGLE complement of the concept of  equilibrium phase diagram. In Section  \ref{section4} we discuss the various density and temperature regimes of the behavior of the long-$t$ asymptotic limit $S^a(k)$ of the NESF $S(k;t)$, whereas in Section \ref{section5}  this analysis is extended to the finite waiting-time regime, to illustrate the predicted scenario of the aging of the structural properties.  In  Section \ref{section6} we summarize the main results of the present paper.

\section{The NE-SCGLE theory.}\label{section2}

The fundamental origin of the 
NE-SCGLE was laid down in detail in Refs. \cite{nescgle1,nescgle2,nescgle3}. However,  a practical summary can be found in the supplementary material of Ref. \cite{nescgle8}, which highlights the main simplifying approximations leading to the version of this theory employed in this and in all its previous concrete applications. The present study will be a straightforward extension of the work reported in Refs.  \cite{olais1,olais2,nescgle8,zepeda}, which describes the predicted  scenario of arrested spinodal decomposition in liquids with excluded-volume plus attractive interactions. The reader is invited to visit these references, which describe in all detail the conceptual and practical challenges found in applying the theory to quenches inside the spinodal region or inside other thermodynamically unstable regions of the state space. These references also illustrate the wealth of information provided by the NE-SCGLE on the time-dependent physical properties of an attractive system under these non-equilibrium conditions. In particular, the novel concept of time-dependent non-equilibrium phase diagram that emerges from these applications is carefully explained in Ref. \cite{zepeda}. 

\subsection{ The NE-SCGLE equations.}\label{subsection2.1}

In its simplest version, the NE-SCGLE theory is summarized by a set of equations that describe the irreversible evolution of the non-equilibrium structural and dynamical properties of a model glass-forming liquid, formed by $N$ identical spherical particles in a volume $V$ that interact through a radially-symmetric pair potential $u(r)$. It starts with the time evolution equation for the  NESF $S(k;t) \equiv \overline{ \delta n({\bf k};t) \, \delta n(-{\bf k};t)}$, where the over-line indicates the average over a (non-equilibrium) statistical ensemble, and where $\delta n({\bf k};t)$ is the Fourier transform of the fluctuations in the local particle number density  $n({\bf r};t)$. For a system that is instantaneously quenched at time $t=0$ from initial bulk density and temperature  $(n_i,T_i)$ to new final values $(n,T)$, constrained to remain spatially uniform ($\overline{n}({\bf r};t)=n\equiv N/V$), such an equation reads, for $t>0$,
\begin{equation}
\frac{\partial S(k;t)}{\partial t} = -2k^2 D^0
b(t)n\mathcal{E}(k;n,T) \left[S(k;t)
-1/n\mathcal{E}(k;n,T)\right], \label{dsktenst}
\end{equation}
with  $D^0$ being the short-time self-diffusion coefficient \cite{shorttimed0}.

In this equation the  time-dependent mobility function $b(t)$ is defined as $b(t)\equiv D_L(t)/D^0$, with $D_L(t)$ being the long-time self-diffusion coefficient at evolution time $t$. This function couples the structural relaxation described by Eq. (\ref{dsktenst}) with the non-equilibrium relaxation of the dynamic properties of the fluid. Such coupling is established by the following exact expression for  $b(t)$,
\begin{equation}
b(t)= [1+\int_0^{\infty} d\tau\Delta{\zeta}^*(\tau; t)]^{-1},
\label{bdtp}
\end{equation}
in terms of the $t$-evolving, $\tau$-dependent friction function $\Delta{\zeta}^*(\tau; t)$, for which the NE-SCGLE theory derives the following approximate expression \cite{nescgle1},
\begin{equation}
\begin{split}
  \Delta \zeta^* (\tau; t)= \frac{D_0}{24 \pi
^{3}n}
 \int d {\bf k}\ k^2 \left[\frac{ S(k;
t)-1}{S(k; t)}\right]^2  \\ \times F(k,\tau; t)F_S(k,\tau; t),
\end{split}
\label{dzdtquench}
\end{equation}
in terms of the NESF $S(k; t)$ and of the NEISF $F(k,\tau; t)\equiv N^{-1}  \overline{ \delta n(\mathbf{k},t+\tau) \delta n(-\mathbf{k},t)}$, where  $\delta n(\mathbf{k},t)$ is the FT of the thermal fluctuations $\delta n(\mathbf{r},t)\equiv n(\mathbf{r},t)-n$ of the local number density  $n(\mathbf{r},t)$ at time $t$. The self-NEISF $F_S(k,\tau; t)$ is defined as $F_S(k,\tau; t)\equiv  \overline{ \exp \left[    i\mathbf{k}\cdot \Delta \mathbf{r}_T(t,\tau)    \right]}$, with $\Delta \mathbf{r}_T(t,\tau) \equiv  \left[ \mathbf{r}_T(t+\tau)-\mathbf{r}_T(t)\right]$ being the displacement of one particle considered as a  tracer. As before, the over-line indicates an average over a corresponding non-equilibrium statistical ensemble.

The previous equations are complemented by the  memory-function equations for  $F(k,\tau; t)$ and $F_S(k,\tau; t)$, written approximately, in terms of their Laplace transforms (LT) $F(k,z; t)$ and $F_S(k,z; t)$, as
\begin{gather}\label{fluctquench}
 F(k,z; t) = \frac{S(k; t)}{z+\frac{k^2D^0 S^{-1}(k;
t)}{1+\lambda (k)\ \Delta \zeta^*(z; t)}},
\end{gather}
and
\begin{gather}\label{fluctsquench}
 F_S(k,z; t) = \frac{1}{z+\frac{k^2D^0 }{1+\lambda (k)\ \Delta
\zeta^*(z; t)}},
\end{gather}
where $\Delta \zeta^*(z; t)$ is the LT of $\Delta \zeta^*(\tau; t)$. In these equations $\lambda (k)\equiv 1/[1+( k/k_{c}) ^{2}]$ is an ``interpolating function" \cite{arrest1}, with $k_{c}$ being an empirically determined cutoff wave-vector. In the present work we use $k_{c}= 1.305(2\pi)/\sigma$, with $\sigma$ being the hard-core particle diameter of our HSDY model fluid, which guarantees that the hard-sphere liquid will have its dynamic arrest transition at a volume fraction $\phi_c=0.582$, in agreement with simulations  \cite{gabriel}. 

Eqs. (\ref{dsktenst})-(\ref{fluctsquench}) constitute the mathematical summary of the  NE-SCGLE theory. Since the function $\mathcal{E}(k;n,T)$ is considered known, they constitute a closed system of equations for the time-dependent SF $S(k;t)$ and for the non-equilibrium dynamic properties  $b(t)$, $\Delta{\zeta}^*(\tau; t)$, $F(k,\tau; t)$, and $F_S(k,\tau; t)$. Thus, Subsection \ref{subsection2.2} defines the thermodynamic function $\mathcal{E}(k;n,T)$, explains its role in determining stability and structural properties.

\subsection{Thermodynamic stability function $\mathcal{E}(k;n,T)$.}\label{subsection2.2}

The function $\mathcal{E}(k;n,T)$ is defined as the FT of $\mathcal{E}[r;n,T]$ which, in turn, is the second functional derivative of the Helmholtz free energy density-functional $\mathcal{F}[n,T]$, 

\begin{equation}
\mathcal{E}[\mid \mathbf{r}-\mathbf{r}'\mid ; n,T] \equiv  \left( \frac{\delta^2 \mathcal{F}[n,T]/k_BT}{\delta n(\mathbf{\mathbf{r}})\delta n(\mathbf{\mathbf{r}'})} \right),
\label{defedk}
\end{equation}
evaluated at the uniform (bulk) density and temperature fields $n(\textbf{r},t)=n\equiv N/V$ and $T(\textbf{r},t)=T$. We refer to  $\mathcal{E}(k;n,T)$ as the thermodynamic stability function, since it provides a criterion for the thermodynamic stability of the system. For example, the state function $\mathcal{E}(k;n,T)$, evaluated at $k=0$, is related with the thermodynamic derivative $ (\partial p/ \partial n)_T$, where $p$ is the pressure, as 
\begin{equation} 
\beta (\partial p/ \partial n)_T = n\mathcal{E}(k=0;n,T),
\label{compressequation}
\end{equation}
which is the so-called compressibility equation \cite{mcquarrie} ($\beta^{-1}\equiv k_BT$). As we see in Subsection \ref{subsection2.3}, this equation of state allows us to determine the main features of the gas-liquid coexistence region, including the spinodal line, obtained from the condition  $\mathcal{E}(k=0;n,T)=0$, which separates the state space $(n,T)$ in the stable ($\mathcal{E}(k=0;n,T)>0$) and the unstable ($\mathcal{E}(k=0;n,T)<0$) equilibrium domains. 

More generally, the function $\mathcal{E}(k;n,T)$ must be positive for all wave-vectors for the system to be stable, since if $\mathcal{E}(k;n,T)<0$ for a finite $k$, the system would be unstable to density fluctuations of wavelength $2\pi/k$.  Under some conditions, a domain $k^- \le k \le k^+$ may exist in which  $\mathcal{E}(k;n,T)<0$. Then, in analogy with the spinodal line, we define a $\lambda$ line in the state space $(n,T)$ by the threshold condition $k^- = k^+ = k_{\lambda}$, at which  $\mathcal{E}(k_{\lambda};n,T)=0$. 

The stability function $\mathcal{E}(k;n,T)$ also plays a relevant role in determining the structural properties of the system. For example, the \emph{equilibrium} static structure factor (SF) $S^{eq}(k)$ is given by 
\begin{equation} 
S^{eq}(k) = 1/n\mathcal{E}(k;n,T)
\label{eqstructurefactor}
\end{equation}
which  is just the well-known Ornstein-Zernike (OZ) equation \cite{mcquarrie}, as can be seen by writing $S^{eq}(k) =1+nh^{eq}(k)$ and $n\mathcal{E}(k;n,T)=1-nc(k)$, with $h^{eq}(k)$ being the FT of the equilibrium \emph{total} correlation function (TCF) $h^{eq}(r)$ and $c(k)$ being the FT of the \emph{direct} correlation function (DCF) $c(r)$. Clearly, the SF $S^{eq}(k)$ of a uniform system does not exist when $\mathcal{E}(k;n,T)<0$.

From the kinetic perspective of Eq. (\ref{dsktenst}), the OZ equation in Eq. (\ref{eqstructurefactor}) is just the equilibrium condition for the NESF $S(k;t)$.  This is, in fact, the obvious \emph{stationary} solution of the kinetic equation for $S(k;t)$ in Eq. (\ref{dsktenst}), asymptotically attained at long times in the kinetic process of equilibration, i.e., $S(k;t\to\infty)=S^{eq}(k) = 1/n\mathcal{E}(k;n,T)$. However, under general non-equilibrium conditions, such as during the transient that follows an instantaneous quench, the NESF $S(k;t)$ is certainly NOT given by the OZ equation $S(k;t)= 1/n\mathcal{E}(k;n,T)$, but by the solution of Eq. (\ref{dsktenst}), which provides a general manner to process the information contained in $\mathcal{E}(k;n,T)$ to determine $S(k;t)$. Thus, one can say that Eq. (\ref{dsktenst}) is the non-equilibrium extension of the OZ equation. Notice that, as a thermodynamic input of this kinetic equation, negative values of $\mathcal{E}(k;n,T)$ are perfectly physical, leading in fact to some of the most remarkable predictions of the NE-SCGLE theory.


\subsection{Mean field free energy functional $\mathcal{F}[n,T]$.}\label{subsection2.3}

In practice, we must identify a pertinent strategy to approximate $\mathcal{F}[n,T]$ for our three-dimensional HSDY model fluid.  
With the aim of adopting the same level of approximation as in the early theoretical discussions of the equilibrium phase diagram of SALR systems, we follow Refs.  \cite{seargelbart,archerpinievansreatto07,archerionescu3Djpcm08} in writing $\mathcal{F}[n,T]$ as the sum $\mathcal{F}[n,T]=\mathcal{F}^{HS}[n,T]+ \mathcal{F}^{DY}[n,T]$ of  the exact hard-sphere free energy $\mathcal{F}^{HS}[n,T]$ plus the contribution $ \mathcal{F}^{DY}[n,T]$ of the double Yukawas. A simple explicit expression for the latter is  provided by its \emph{random phase approximation} (RPA), within which  $\mathcal{F}[n,T]$ is given by
\begin{equation}
\mathcal{F}[n,T]=\mathcal{F}^{HS}[n,T]+\frac{1}{2} \int d\mathbf{r} \int d\mathbf{r}' n(\textbf{r})u^{DY}(\mid\mathbf{r}-\mathbf{r}'\mid) n(\textbf{r}'), 
\label{feqfhsplusfdy1}
\end{equation}
and the function $\mathcal{E}[\mid\mathbf{r}-\mathbf{r}'\mid ; n,T]$ of Eq. (\ref{defedk}) is given by
\begin{equation}
\mathcal{E}[\mid\mathbf{r}-\mathbf{r}'\mid ; n,T]=\mathcal{E}^{HS}[\mid\mathbf{r}-\mathbf{r}'\mid ; \phi]+\beta u^{DY}(\mid\mathbf{r}-\mathbf{r}'\mid), 
\label{feqfhsplusfdy2}
\end{equation}
whose FT reads
\begin{equation}
\mathcal{E}(k; n,T)=\mathcal{E}^{HS}(k; \phi)+\beta u^{DY}(k).
\label{feqfhsplusfdy3}
\end{equation}

Writing $\mathcal{F}^{HS}[n,T]$ as $\mathcal{F}^{HS}[n,T]=\mathcal{F}_{id}^{HS}[n,T] + \mathcal{F}^{HS}_{ex}[n,T]$, where $\mathcal{F}_{id}^{HS}[n,T]$ is the ideal-gas value  and $\mathcal{F}_{ex}^{HS}[n,T]$ is the ``excess'' HS contribution \cite{evans}, allows us to write Eq. (\ref{defedk}) for the pure HS system as  
\begin{equation}
\mathcal{E}^{HS}[\mid\mathbf{r}-\mathbf{r}'\mid ; n,T] = \delta (\mathbf{r}-\mathbf{r}')/n-c^{HS}(\mid\mathbf{r}-\mathbf{r}'\mid ; n,T)
\label{feqfhsplusfdy4}
\end{equation}
with $c^{HS}(\mid\mathbf{r}-\mathbf{r}'\mid ; n,T)$, defined by $c^{HS}(\mid\mathbf{r}-\mathbf{r}'\mid ; n,T)\equiv -\beta \left( \delta^2 \mathcal{F}_{ex}^{HS}[n,T]/\delta n(\mathbf{\mathbf{r}})\delta n(\mathbf{\mathbf{r}'}) \right)$, being the exact HS DCF. In Fourier space this equation reads  $\mathcal{E}^{HS}(k; \phi)=1/n-c^{HS}(k; \phi)$. As in Refs.  \cite{archerpinievansreatto07,archerionescu3Djpcm08}, here we will also approximate $c^{HS}(k; \phi)$ by its Percus-Yevick (PY) approximation \cite{percusyevick}, but complemented with its Verlet-Weis (VW) correction \cite{verletweis}, $c^{HS}(k; \phi) \approx c^{HS}_{PYVW}(k; \phi)$. Thus, the approximate thermodynamic input $\mathcal{E}(k;n,T)$ that we shall employ in this work is finally written as 
\begin{equation} 
\mathcal{E}(k;n,T) \approx 1/n-c^{HS}_{PYVW}(k; \phi) + \beta u^{DY}(k).
\label{approxedk}
\end{equation}

\begin{figure*}[ht!]
\subfigure{\label{Fig:1a}
\includegraphics[scale=0.2]{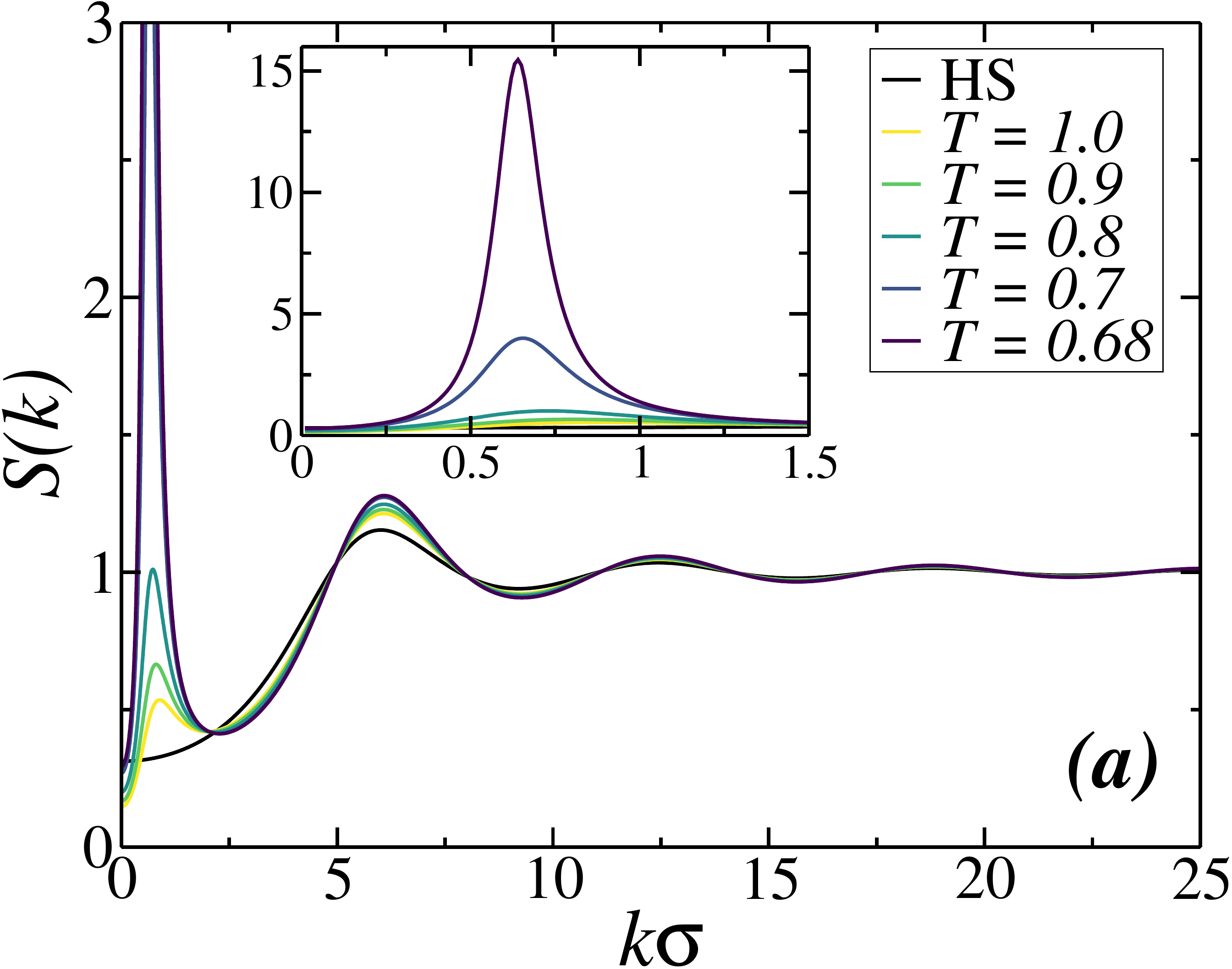}
}
\subfigure{\label{Fig:1b}
\includegraphics[scale=0.2]{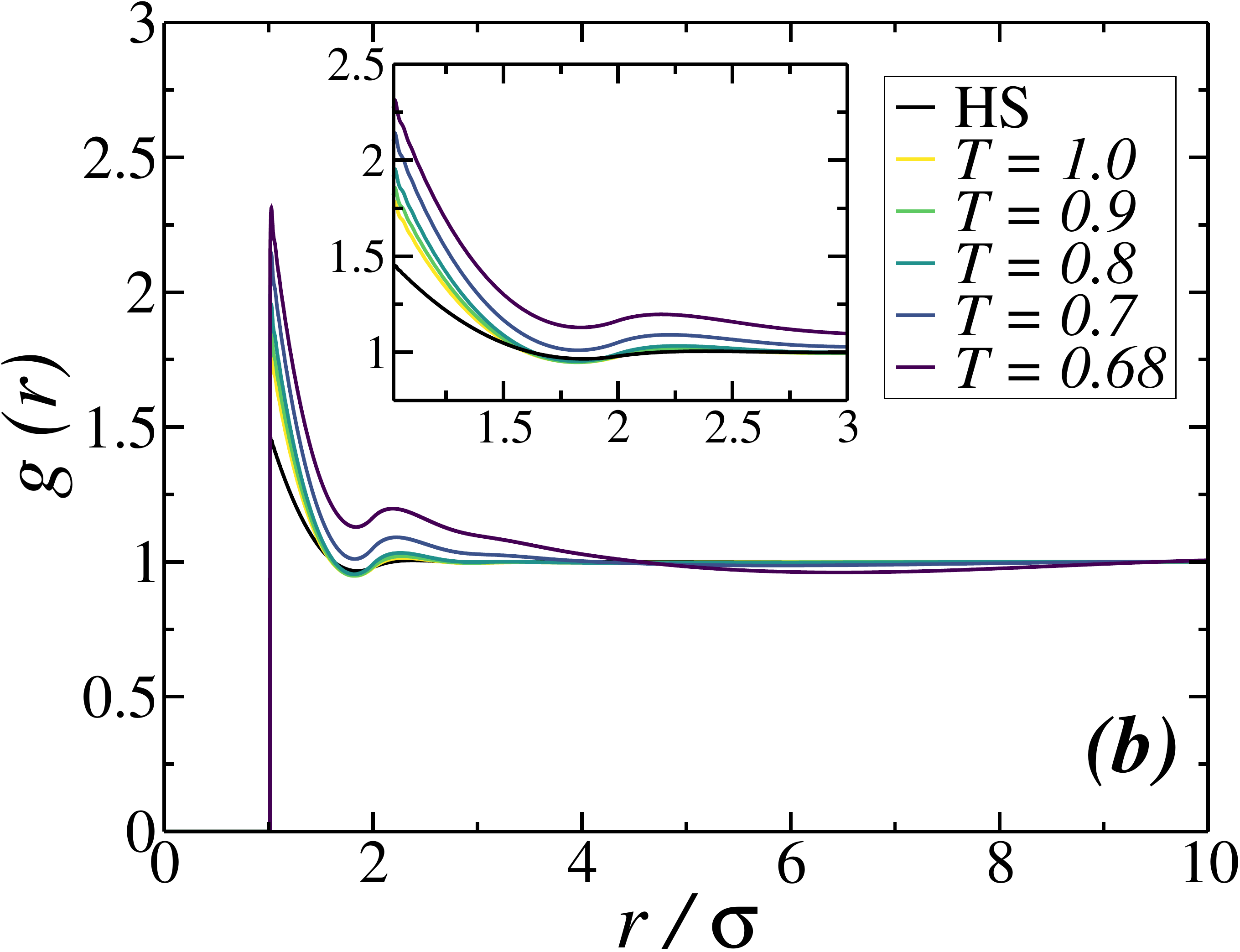}
}
\subfigure{\label{Fig:1c}
\includegraphics[scale=0.2]{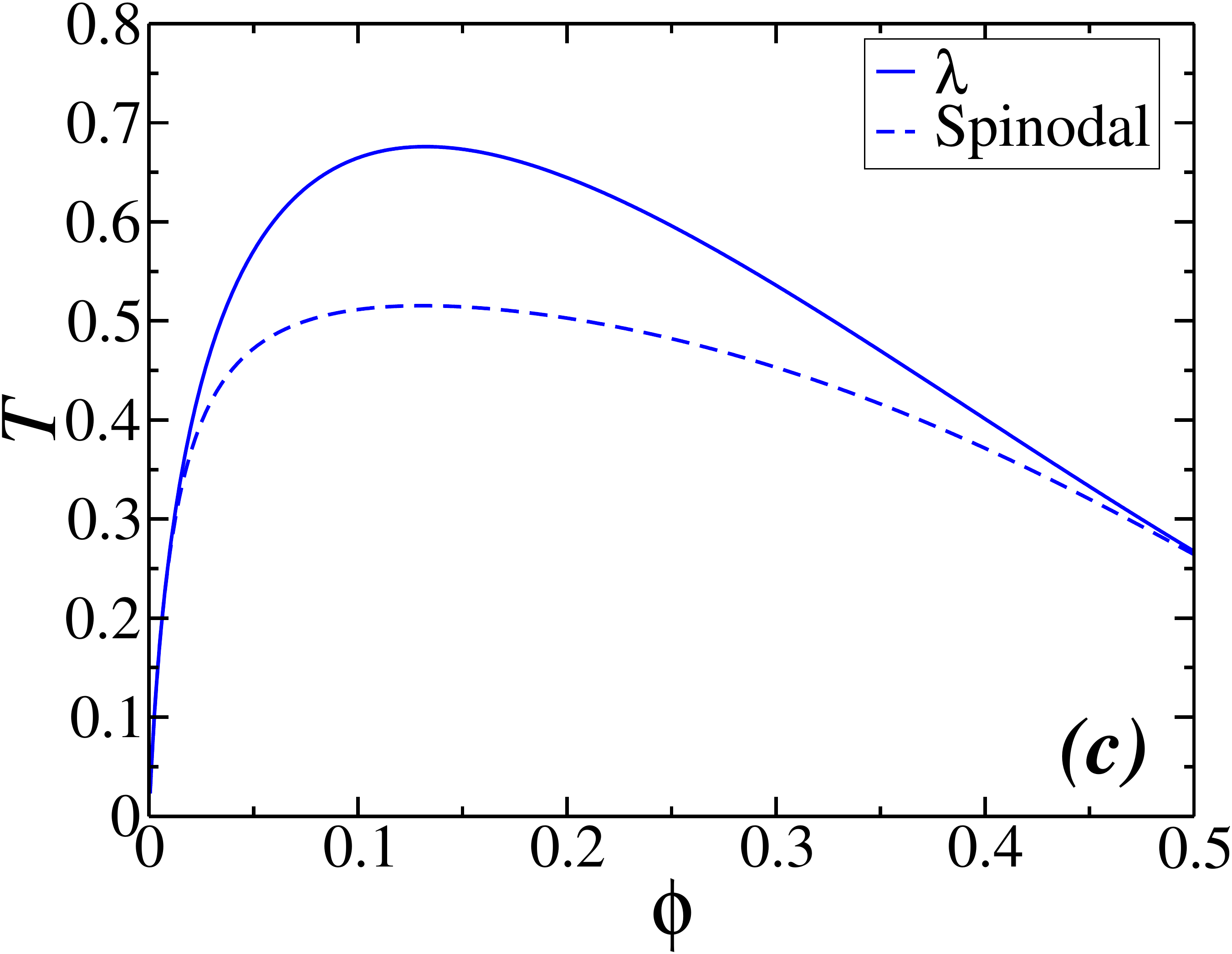}
}
\caption{Behavior of (a) the SF $S(k;n,T) = 1/n\mathcal{E}(k;n,T)$ and (b) the corresponding RDF $g(r;\phi,T)$, as provided by the RPA (Eq. (\ref{approxedk})) for the HSDY fluid, with parameters $z_1=1$, $z_2=0.5$ and $A=0.5$, for fixed volume fraction $\phi=0.15$ and final temperatures above the 
$\lambda$ line, $T>T_\lambda (\phi) =0.6733$, as indicated. 
(c) Equilibrium  gas-liquid  phase diagram of the same HSDY fluid, exhibiting the spinodal (dashed line) and the  $\lambda$ (solid line)  curves. }
\label{fig1}
\end{figure*}

\subsection{Equilibrium properties of the HSDY model.}\label{section2.4}

We may now use this approximate expression for the thermodynamic input $\mathcal{E}(k;n,T)$  of the NE-SCGLE equations, Eqs. (\ref{dsktenst})-(\ref{fluctsquench}), whose solution will describe the non-equilibrium response of the liquid after the instantaneous quench. For reference, however, let us first summarize some purely equilibrium properties that directly derive from  the RPA  for the thermodynamic input $\mathcal{E}(k;n,T)$ in Eq.  (\ref{approxedk}), applied to the HSDY  potential. For this, let us rewrite the DY term in Eq. (\ref{hsdy})  in dimensionless form as 
\begin{equation}
\beta u^{DY}(r) = -\epsilon \frac{\exp[-z_1(r/\sigma-1)]}{r/\sigma}+ A \frac{\exp[-z_2(r/\sigma-1)]}{r/\sigma}.
\label{hsdy1}
\end{equation}
with $\epsilon \equiv \epsilon_1/k_BT$ and $A\equiv \epsilon_2/k_BT$. 

Following Refs. \cite{archerpinievansreatto07,archerionescu3Djpcm08}, we have in mind a colloid-polymer mixture, and consider the (``athermal'') conditions in which the physical temperature $T$ is kept fixed and the strength of the attraction is controlled by varying the polymer concentration, i.e., $\epsilon^{-1}$ plays the role of an effective temperature $T_{eff}=\epsilon^{-1}$, which for simplicity in notation will be denoted as $T$. Thus, if the particle charge and ionic strength can be assumed fixed, we can treat the parameters  $A$ and $z_2$ (the intensity and range of the electrostatic repulsion) as fixed parameters. As a result, the function $\mathcal{E}(k;n,T)$ will actually depend on the dimensionless concentration $n^*\equiv n \sigma^3$, which we shall continue denoting by $n$ (or in terms of the packing fraction $\phi=\pi n /6$), and on the effective temperature T. This dependence is additional to the dependence of  $\mathcal{E}(k;\phi,T)$ on the potential parameters $A$, $z_1$ and $z_2$.

\vspace{-15pt}
\subsection{Equilibrium structure of the HSDY model.}\label{subsection2.5}

The first relevant property to discuss is, of course, the SF $S^{eq}(k;n,T)$ given by Eq. (\ref{eqstructurefactor}), which for notational convenience we will denote as $S(k;n,T)$. From now on, we will consider a HSDY fluid with fixed parameters $z_1=1$, $z_2=0.5$ and $A=0.5$, which according to  Ref. \cite{LiuXi2019}, corresponds to a type III SALR system. Fig. \ref{Fig:1a} exhibits the main distinctive feature of the SF, previously discussed by Sear and Gelbart \cite{seargelbart} and by Archer et al.  \cite{archerpinievansreatto07} for the same HSDY model within the RPA. We refer to the development of a large peak centered at a small but finite wave vector $k = k_\lambda$ (referred to as the ``cluster'' peak \cite{archerpinievansreatto07}). This peak is  associated with the propensity to the spontaneous formation of clusters in the equilibrium fluid, with its height $S(k_\lambda;\phi,T)$ increasing, and its position $k_\lambda$ decreasing, as the temperature $T$ is reduced along a given isochore ($\phi=0.15$, in the present case).

According to the RPA results in Fig. \ref{Fig:1a},  $S(k_\lambda;\phi,T)$ actually diverges as the temperature $T$ reaches a singular value $T_\lambda (\phi)$ from above. The curve $T=T_\lambda (\phi)$, defined by the condition $\mathcal{E}(k_\lambda;\phi,T)=1/S(k_\lambda;\phi,T)=0$, is referred to as the \emph{$\lambda$ line} (for the isochore of the figure, $T_{\lambda}(\phi=0.15)=0.6733$)). Taking the inverse FT of $[S(k)-1]/n$ yields the total correlation function $h(r)$, and in Fig. \ref{Fig:1b} we plot the same information of Fig. \ref{Fig:1a}, but in terms of the radial distribution function $g(r) = 1+h(r)$. The results in this figure exhibit a notorious structural feature as $T$ approaches $T_\lambda (\phi)$ from above, namely the emergence of a long-range minimum in $g(r;\phi,T)$ at $r=r_{min}$ ($\approx 6\sigma$ in the RDF corresponding to the deepest quench, $T=0.68$,  in the figure). As we shall see later, this structural feature is an essential fingerprint of the process of cluster formation and eventual dynamic arrest due to cluster-cluster repulsions. Finally, notice that Figs. \ref{Fig:1a} and \ref{Fig:1b}  report the same information as Figs. 1 and 2 of  Archer et al. \cite{archerpinievansreatto07}, although at a different value of the parameter $A$ (=\ 0.5 here, and =\ 0.082 in Ref. \cite{archerpinievansreatto07}). 

At this point it is pertinent to clarify that the appearance of a low-$k$ peak in the SF is a characteristic hallmark of many SALR systems (whether type I, II and III) and it is generally associated with the emergence of additional ordering at a length scale much larger than the particle diameter $\sigma$. As discussed in Ref. \cite{liuJPCB}, particularly referring to  lysozyme protein solutions, this low-$k$ peak is not necessarily related to the formation of clusters with optimal sizes and hence, should more appropriately be referred to as an  \emph{intermediate range order} (IRO)  peak \cite{liuJPCB}. Notwithstanding this clarification, we can say, following the discussion of Refs. \cite{archerpinievansreatto07,archerwilding}, that in our present type III SALR model system, it is safe to refer to $S(k_{\lambda};\phi,T)$ as a ``cluster" peak.


\subsection{Equilibrium phase diagram of the HSDY model.}\label{subsection3.2}
\vspace*{-5pt}
The $\lambda$ line $T_{\lambda}(\phi)$ of the system described in Figs. \ref{Fig:1a} and \ref{Fig:1b} is represented in Fig. \ref{Fig:1c}. For reference, the spinodal line is also plotted
and it was calculated from the approximate function $\mathcal{E}(k;n,T)$  in Eq. (\ref{approxedk}). This figure, which is our version of Fig. 6 of Ref.  \cite{archerpinievansreatto07}, summarizes the equilibrium phase diagram of the HSDY model predicted by the RPA, which indicates that the gas-liquid transition is preempted by the occurrence of the singular $\lambda$ line, below which, uniform disordered equilibrium states do not exist, since $S(k;\phi,T)=1/n\mathcal{E}(k;\phi,T)$ is negative (and hence, non-physical) for wave-vectors in an interval around $k_\lambda$.

With the aim of understanding this puzzling feature of the equilibrium phase behavior of the type III  HSDY model, Archer and Wilding \cite{archerwilding} carried out Monte Carlo simulations, showing that, under some circumstances, the repulsive part of the pair potential could lead to the replacement of the liquid-vapor coexistence by two first-order phase transitions. The first involves the coexistence of the vapor with a fluid of spherical liquid-like clusters, and the second involves the coexistence of a phase of spherical voids with the homogeneous liquid. These two transition lines meet with the vapor-liquid transition at a triple point. Later on, using density-functional theory, Archer et al. \cite{archerionescu3Djpcm08} concluded that  below the $\lambda$ line, the equilibrium states consisted of non-uniform spatially-ordered (or ``modulated'') phases (see Fig. 5 of Ref. \cite{archerionescu3Djpcm08}), thus theoretically supporting in part the equilibrium scenario advanced by Monte Carlo calculations.

\section{Glass transition diagrams.}\label{section3}
\vspace*{-5pt}

Unfortunately, careful experimental work in colloidal systems with competing interactions does not report the observation of these equilibrium ordered structures \cite{sedgwick,stradner,bordi,campbell,DibbleKoganSolomonPRE06,KlixRoyalTanakaPRL2010}. Instead, as also suggested by extensive molecular dynamics simulations \cite{sciortinomossa,mossasciortino,sciortinotartaglia}, what seems to surround or substitute the predicted spatially-ordered phases are equilibrium fluids of finite-sized clusters and percolating gel-like states.
So far, the experimental elusiveness in observing the aforementioned modulated phases in SALR systems remains unresolved. One possible reason, of course, is the fact such phases are theoretically predicted for type III SALR liquids while -- according to Ref. \cite{LiuXi2019} -- most model experimental SALR systems studied are of type I. A second possibility, however, is related to the fact that the cluster fluid and percolated gel-like states are able to evolve into genuinely arrested non-equilibrium states, whose nature is controlled by the interplay between repulsions and attractions, and whose fundamental understanding poses even more complex technical and fundamental challenges. 

The theoretical description of such dynamically-arrested states, in fact, is typically out of the scope of conventional equilibrium theories and simulations, and hence, must be described from a genuinely non-equilibrium perspective, such as that provided by  the solution of  the  NE-SCGLE equations (\ref{dsktenst})-(\ref{fluctsquench}). In analyzing the solution of this set of equations, it is instructive to start by discussing their asymptotic long-time stationary limits. For this, we can consider two strategies: either we first take the stationary limit of Eqs.  (\ref{dsktenst})-(\ref{fluctsquench}) and then analyze its solutions, or else, we first formally determine the full time-dependent solution of Eqs.  (\ref{dsktenst})-(\ref{fluctsquench}) and then take the long-time stationary limit. Let us describe the results of these two strategies.

\vspace*{-5pt}
\begin{figure*}{}
\begin{center}
\subfigure{\label{Fig2a}
\includegraphics[scale=0.95]{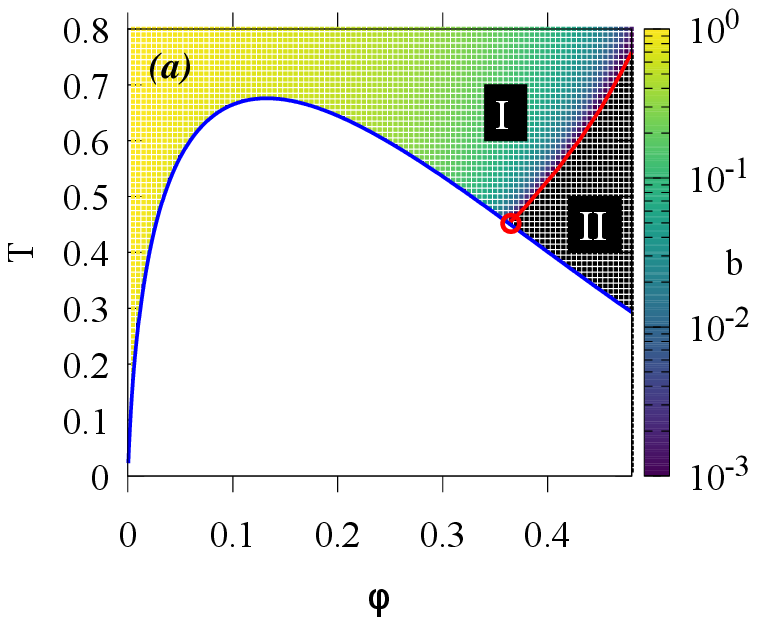}
}
\hskip.4cm
\subfigure{\label{Fig2b}
\includegraphics[scale=0.95]{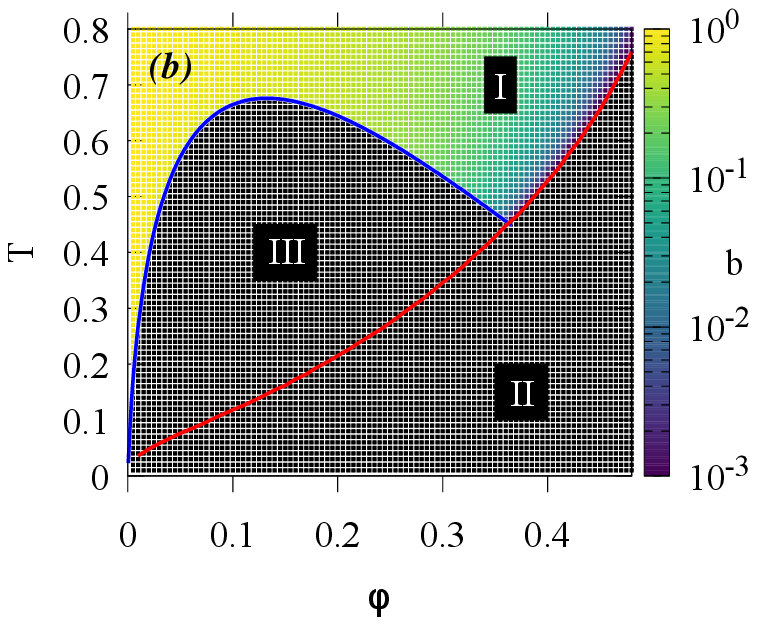} 
}
\caption{(a) \emph{Glass transition} diagram of the HSDY fluid with $z_1=1$, $z_2=0.5$ and $A=0.5$, provided by the equilibrium SCGLE theory using Eq. (\ref{gammaeq}). The blue solid line is the $\lambda$ line, and the red solid one is the fluid-glass transition line, with both meeting at the (``bifurcation'') point $(\phi_b,T_b)=(0.365,0.452)$ (open circle). Region I identifies the ergodic region and region II identifies the glass region.  The predicted equilibrium mobility at each state points $(\phi,T)$ is represented by a color code (colored region) above the $\lambda$ line, with the black color representing the limiting condition $b^{eq}(\phi,T)=0$. In the light empty region below the $\lambda$ line, Eq. (\ref{gammaeq}) cannot be used since $S(k;\phi,T)$ is non-physical. (b) Non-equilibrium glass transition diagram provided by  the NE-SCGLE theory using Eq. (\ref{nep5ppdu}). Above the $\lambda$ line, it coincides with the glass transition diagram in (a), but adds the information that the $\lambda$ line itself is a ``type B'' ergodic-to-non-ergodic transition line, and that the fluid-glass transition line continues from above to below the $\lambda$ line as a ``type B'' non-ergodic to non-ergodic transition line. }
\label{Fig2}
\end{center}
\end{figure*}
\subsection{MCT-like ``glass transition'' diagram.}\label{subsection3.3}
\vspace*{-5pt}
The first strategy only considers stationary solutions that correspond to thermodynamic equilibrium states, and only requires the knowledge of the thermodynamic function $\mathcal{E}(k;n,T)$. Its main result  is the MCT-like ``glass transition'' diagram of the system \cite{zepeda,sperl1}, which determines the borderline of the region of the state space where no kinetic barriers will impede the system from reaching thermodynamic equilibrium.


To explain the concept of  \emph{glass transition diagram} (GTD), let us notice that  Eqs. (\ref{dsktenst})-(\ref{fluctsquench}) contain the description of the dynamic properties of \emph{equilibrium fluids} as the particular limit in which the solution $S(k;t)$ of Eq. (\ref{dsktenst}) has reached its equilibrium limit $S(k;t\to\infty)=S(k;n,T) = 1/n\mathcal{E}(k;n,T)$. In this limit one recovers the original \emph{equilibrium} SCGLE theory of dynamical arrest \cite{arrest1,arrest2,arrest3}, that describes the dynamics of \emph{equilibrium} fluids. 
The corresponding SCGLE equations follow from replacing $S(k;t)$,  $\Delta{\zeta}^*(\tau; t)$, $F(k,\tau; t)$, and $F_S(k,\tau; t)$ in Eqs. (\ref{dzdtquench})-(\ref{fluctsquench}) by, respectively,  $S(k)$, $\Delta{\zeta}^{*eq}(\tau)\equiv \Delta{\zeta}^*(\tau; t\to \infty)$, $F^{eq}(k,\tau)\equiv F(k;\tau; t\to \infty)$ and $F_S^{eq}(k,\tau)\equiv F_S(k;\tau; t\to \infty)$. As in MCT, the SCGLE theory provides equations for the so-called non-ergodicity parameters, which are the long-$\tau$ asymptotic limits $f^{eq}(k) \equiv F^{eq}(k,\tau\to \infty)$, and $f^{eq}_S(k)\equiv F_S^{eq}(k,\tau\to \infty)$.

In the case of the SCGLE theory, $f^{eq}(k)$ and $f_S^{eq}(k)$ can be written in terms of the SF, $S(k;n,T)$, and the squared localization length $\gamma^{eq}(\phi,T)$ (Eqs. (9) and (10) of Ref. \cite{scgle3}), the latter being solution of the equation 
\begin{equation}
\begin{split}
\frac{1}{\gamma^{eq}(\phi,T)} =
\frac{1}{6\pi^{2}n}\int_{0}^{\infty }
dkk^4\left[S (k;n,T)-1\right] ^{2}\lambda^2(k)\times \\
\frac{1}{\left[\lambda (k)S (k;n,T) +
k^2\gamma^{eq}(\phi,T)\right]\left[\lambda (k) + k^2\gamma^{eq}(\phi,T)\right]}.
\end{split}
\label{gammaeq}
\end{equation}
The dynamic order parameter $\gamma^{eq}(\phi,T)$ diverges at equilibrium, and has a finite value at non-ergodic states. Thus, it partitions the state space $(\phi,T)$ into its ergodic and non-ergodic regions, with the ideal glass transition line  being the borderline between them. The monotonically increasing (red) solid line  of Fig. \ref{Fig2a} is such fluid-glass transition line $T=T_c(\phi)$  for our HSDY  model, obtained within  the RPA for $S(k;n,T)$ in Eqs.  (\ref{approxedk}) and (\ref{eqstructurefactor}). This liquid-glass transition line originates at high-temperatures and high-densities at the well-known hard-sphere glass transition, occurring at $\phi_c\approx 0.58$. At lower $T$ and $\phi$, it meets the $\lambda$ line at the state point $(\phi_b,T_b)=(0.365,0.452)$.

Upon crossing this transition line from the ergodic domain (region I  of Fig. \ref{Fig2a}) to the non-ergodic domain (region II), the order parameter $\gamma^{eq}(\phi,T)$ changes discontinuously from its equilibrium infinite value to a finite non-equilibrium value, and this discontinuous behavior characterizes what is referred to (borrowing MCT language) as a ``type B'' glass transition. An alternative dynamic order parameter is the normalized equilibrium long-time self-diffusion coefficient $D_L(\phi,T)/D_0= b^{eq}(\phi,T)= b(t\to\infty;\phi,T)$, also provided by the solution of the SCGLE equations. This dynamic state function is finite and positive in region I and vanishes in region II and along the ideal fluid-glass transition line $T=T_c(\phi)$. Its value is represented in Fig. \ref{Fig2a} by a color code, with the darkest color corresponding to the arrested region II. 

Clearly, since $S(k;n,T)$ does not exist below the  $\lambda$ line (non-monotonic (blue) solid line in Fig. \ref{Fig2a}), the SCGLE theory cannot provide any predictions there. Thus, this method can identify the ergodic and non-ergodic regions only above the  $\lambda$ line, leaving the (light) region below, empty of information. This limitation is, of course, shared by MCT. In fact, our glass transition diagram in Fig. \ref{Fig2a} is the analog of  the MCT ``kinetic phase diagrams'' reported in Fig. 2 of Ref. \cite{wuliuchencao} for different values of  $A$, $z_1$ and $z_2$.  


\subsection{Non-equilibrium ``glass transition'' diagram (NEGTD).}\label{subsection3.4}

The second strategy provides a route of escape from the limitation of the SCGLE theory to equilibrium conditions. For this we must return to the full set of equations (\ref{dsktenst})-(\ref{fluctsquench}) that summarize the more general NE-SCGLE formalism. Following the approach explained in detail in Ref. \cite{olais1}, one can extend the method based on the solution of Eq. (\ref{gammaeq}) for $\gamma^{eq}(\phi,T)$, by introducing the more general parameter $\gamma^*(u)$, 
defined in Ref. \cite{olais1} as the solution of its Eq. (3.6), namely, 
\begin{equation}
\begin{split}
\frac{1}{\gamma^*(u)} =
\frac{1}{6\pi^{2}n}\int_{0}^{\infty }
dkk^4\left[S^*(k;u)-1\right] ^{2}\lambda^2
(k;u)\times \\
\frac{1}{\left[\lambda (k;u)S^*(k;u) +
k^2\gamma^*(u)\right]\left[\lambda (k;u) + k^2\gamma^*(u)\right]},
\end{split}
\label{nep5ppdu}
\end{equation}
where $S^*(k;u)$ is given by
\begin{equation}
\begin{split}
S^*(k;u) \equiv
[n\mathcal{E}(k;\phi,T)]^{-1}+\left \{ S_i(k)-
[n\mathcal{E}(k;\phi,T)]^{-1}\right \}\times \\
e^{-2k^2D_0
n\mathcal{E}(k;\phi,T)u}, 
\end{split}
\label{solsigmadku}
\end{equation}
with $S_i(k)=S^*(k;u=0)$ being an (arbitrary) initial condition. 

\begin{figure*}[ht]
\subfigure{\label{Fig3a}
\includegraphics[scale=0.27]{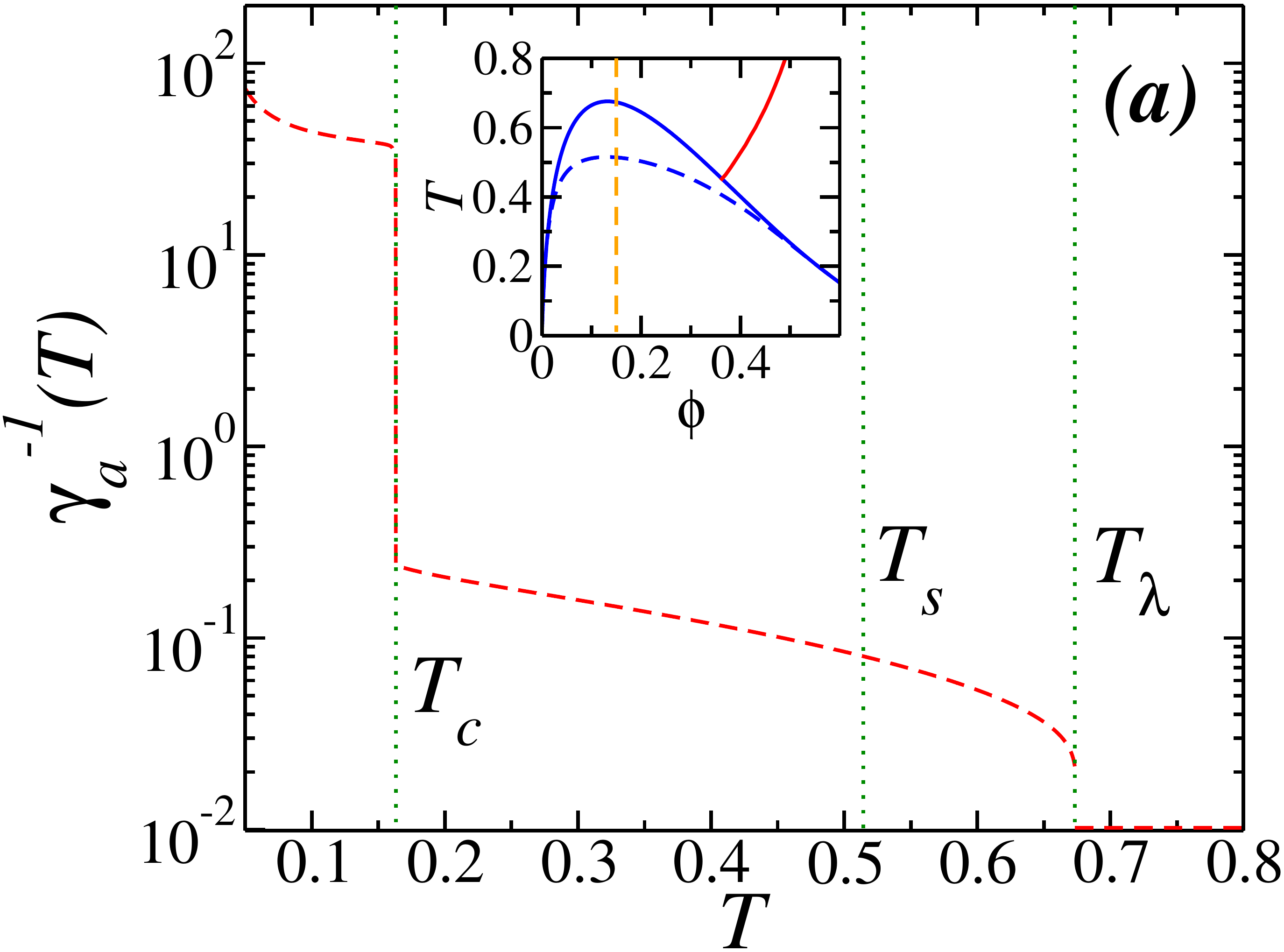}
}
\hskip1cm
\subfigure{\label{Fig3b}
\includegraphics[scale=0.27]{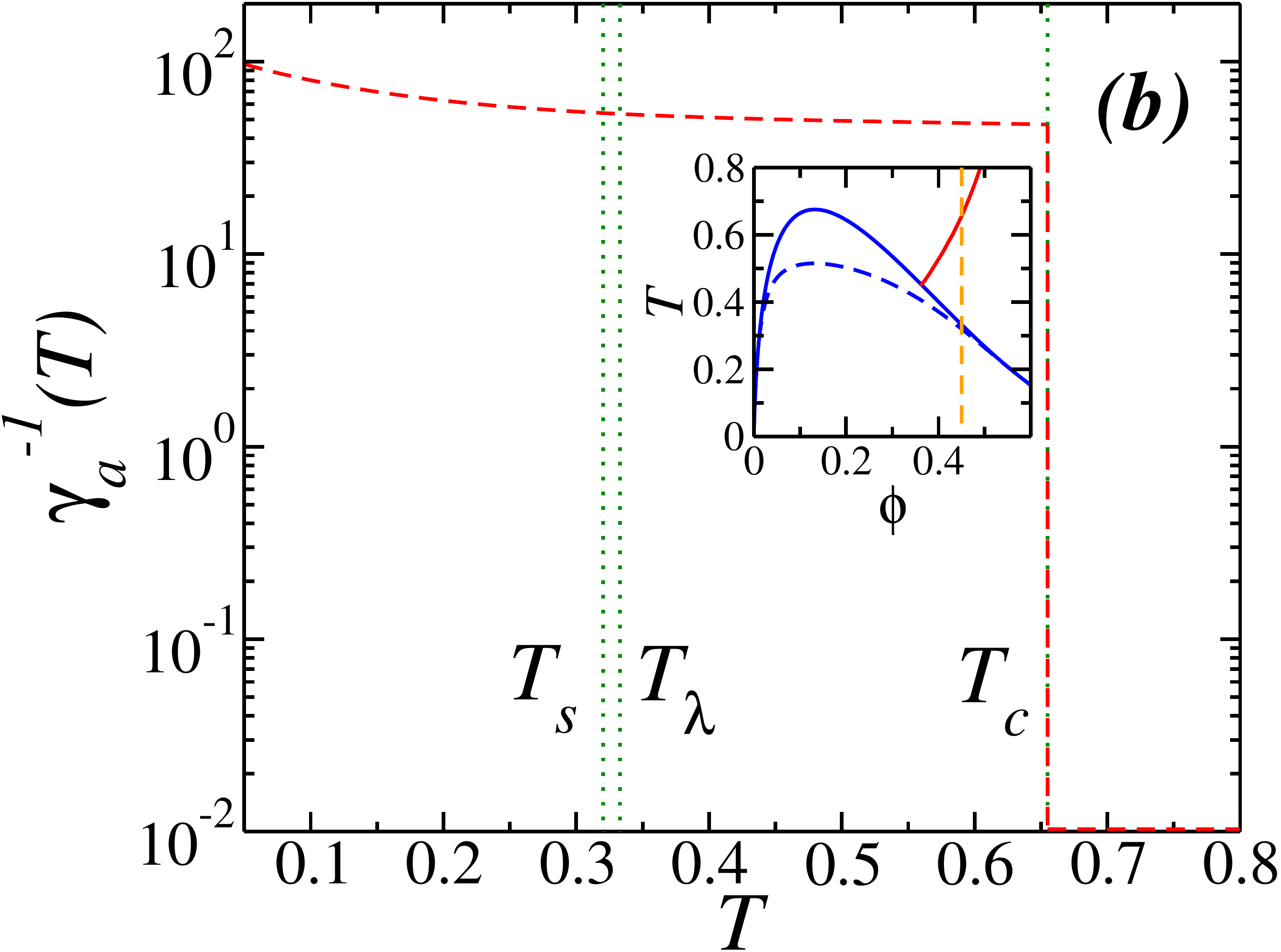}
}
\caption{Dependence of the non-equilibrium dynamic order parameter $\gamma_a(\phi,T)$ on the final temperature $T$, for a sequence of quenches of the HSDY fluid ($z_1=1$, $z_2=0.5$ and $A=0.5$) at fixed volume fraction $\phi$ and with initial condition $S_i(k)=S^{eq}_{HS}(k;\phi)$ for, (a) $\phi=0.15$, to the left of the bifurcation point  $(\phi_b,T_b)=(0.365,0.452)$ (see the vertical dashed line in the inset), and (b) $\phi=0.45$, to the right of the bifurcation point. In (a), the vertical dotted lines indicate the three temperatures corresponding to the $\lambda$ and spinodal lines, and the glass-to-glass transition, thus highlighting the discontinuous jumps of $\gamma_a$ at the two temperatures $T_{\lambda}$ and $T_c$, and also the fact that $T_s$ passes completely unnoticed for the non-ergodicity parameter. In (b), the vertical lines emphasize that $\gamma_a$ only jumps discontinuously at $T_c$, and crosses the two lines of thermodynamic instabilities without noticing them.} \label{Fig3}
\end{figure*}

For a given state point $(\phi,T)$ we use $\mathcal{E}(k;\phi,T)$ to
solve Eq. (\ref{nep5ppdu}), which determines $\gamma^*(u)$ as a function of $u$. If we find that  a finite value $u_a$ of the parameter $u$ exists, such that $\gamma^*(u)$ is infinite  in the finite interval $0\le u < u_a$ and finite for $u_a \le u \le \infty$, then we conclude that the system will become kinetically arrested. If, instead, $\gamma^*(u)=\infty$ for $0\le u \le \infty$, we conclude that the state point $(\phi,T)$ lies in the ergodic region of state space. Thus, the functions $u_a(\phi,T)$ and   $\gamma_a(\phi,T)\equiv \gamma^*(u_a(\phi,T))$, are dynamic order parameters, which allows us to draw what we refer to as the \emph{non-equilibrium glass transition diagram} (NEGTD). In Fig. \ref{Fig2b} we present the results corresponding to our specific HSDY model. Note that, contrary to the algorithm based on Eq. (\ref{gammaeq}), whose input is the SF $S (k;n,T)$ (non-physical below the $\lambda$  line), the input of this non-equilibrium algorithm is the thermodynamic stability function $\mathcal{E}(k;\phi,T)$, whose negative values simply indicate conditions of thermodynamic instability. As a result, this more general criterion to determine dynamic arrest is far more powerful than that  based on Eq. (\ref{gammaeq}), since it is also applicable to quenches to thermodynamically unstable regions of state space, as illustrated by the results in  Fig. \ref{Fig2b}.

The NEGTD of Fig. \ref{Fig2b} completely agrees with the GTD of Fig. \ref{Fig2a}, in that both algorithms partition the portion of  the plane $(\phi,T)$ \emph{above} the $\lambda$ line, into the two well-defined regions: region I of ergodic fluid states and region II of non-ergodic glass states.  However, in the complementary portion of state space, i.e., at and below the  $\lambda$ line,  the NE-SCGLE theory reveals a remarkable non-equilibrium  scenario, whose first unexpected feature is the existence of  two new dynamic arrest transitions for volume fractions below that of the meeting point $(\phi_b,T_b)$. The first of them occurs at a dynamic arrest temperature   that virtually coincides with the $\lambda$ temperature $T_\lambda(\phi)$, and constitutes the boundary between the (ergodic) region I and a new region (region III) of dynamically arrested states, whose nature is expected to be revealed by their non-equilibrium structural and dynamical properties, provided by the full solution of the NE-SCGLE equations.

The second is a dynamic arrest transition occurring along the temperature $T=T_c(\phi)\ (< T_\lambda(\phi))$, which defines the boundary  between the arrested states in region III and the glass states in the low-temperature--low-density extension of region II. These two arrest lines merge at the meeting point $(\phi_b,T_b)$, which is now revealed to actually be a point of bifurcation of the fluid-to-glass transition line of Fig. \ref{Fig2a}. The latter clearly appears to continue below the $\lambda$ line (i.e., for $\phi \le \phi_b$) as the arrested-glass transition line $T=T_c(\phi)$.  

\subsection{NE-SCGLE dynamic order parameter $\gamma_a^*(\phi,T)$.}\label{subsection3.5}

Most of the previous conclusions directly derive from the density and temperature dependence of the dynamic order parameters  $u_a(\phi,T)$ and   $\gamma_a(\phi,T)$ which, in turn, depend on $(\phi,T)$ through the function $\mathcal{E}(k;\phi,T)$ in Eq. (\ref{solsigmadku}). 
The parameter $\gamma_a(\phi,T)$ bears a more transparent physical significance, better explained by its original \emph{equilibrium} definition: it is the long-time asymptotic value $\gamma^{eq} \equiv \lim_{\tau \to \infty} W^{eq}(\tau) $ of the mean squared displacement (MSD) $W^{eq}(\tau)\equiv \langle(\Delta\mathbf{ r}(\tau))^2/3\rangle^{eq}$, of  individual particles. This parameter is infinite if the particles diffuse, and is finite if they are immobilized, in which case it defines the square localization length. This definition extends to non-equilibrium conditions by explicitly including  the waiting time $t$ in the definition of the MSD, which is now defined as $W(\tau;t)\equiv \langle(\mathbf{ r}(t+\tau)-\mathbf{ r}(t))^2/3\rangle$, thus allowing us to define the non-equilibrium square localization length $\gamma_a$ as $\gamma_a \equiv \lim_{\tau \to \infty}\lim_{t \to \infty} W(\tau;t)$.

The use of the function $\gamma_a(\phi,T)$  is better illustrated in Fig. \ref{Fig3}, which plots the inverse of $\gamma_a(\phi,T)$ as a function of the final temperature $T$ along two isochores, $\phi=0.15$ (Fig. \ref{Fig3a}) and $\phi=0.45$  (Fig. \ref{Fig3b}).  
In Fig. \ref{Fig3a}, the (red) dashed horizontal line on the right ($\gamma_a^{-1}(\phi,T)=10^{-2}$) indicates, in reality, that  the function $\gamma_a (\phi=0.15,T)$ actually remains infinite for all final temperatures above a critical temperature of about $0.673$ (which, within the resolution of the figure, coincides with the temperature of the $\lambda$ line for that isochore, $T_\lambda(\phi=0.15)=0.6733$). At this singular temperature,  $\gamma_a(\phi,T)$ jumps \emph{discontinuously} to a finite value $\gamma_a(T_\lambda^-)\approx 40$ (localization length $\sqrt{\gamma_a} \approx 6.3$), remaining finite for $T < T_\lambda(\phi)$. This implies that the $\lambda$ line, besides being the threshold of the thermodynamic stability of uniform equilibrium states (leading to the conjectured equilibrium modulated phases below the $\lambda$ line),  also seems to be an ergodic to non-ergodic transition line. In fact, since $\gamma_a(\phi,T)$  jumps discontinuously at $T =T_\lambda(\phi)$, it is a ``type B'' singularity (in MCT terminology). 

Examining  now temperatures below the $\lambda$ line, we see that the parameter $\gamma_a (T)$ decreases continuously, and exhibits a second discontinuity at a lower temperature $T_c(\phi=0.15)=0.165$. This discontinuity implies the existence of still a second dynamic arrest transition, now corresponding to a non-ergodic--to--non-ergodic (or ``glass-glass'') ``type B'' transition, in which the dynamic order parameter $\gamma_a (T)$  changes discontinuously by about one order of magnitude, from a value $\gamma_a(T_c^+)\approx 4$ to another finite value $\gamma_a(T_c^-)\approx 0.025$ (localization length $\sqrt{\gamma_a} \approx 0.15$, typical of hard-sphere glasses). Besides these two discontinuities, we could not identify any other dynamically singular temperature. In fact, we found that the function  $\gamma_a(\phi,T)$ is perfectly continuous through the spinodal curve $T=T_s(\phi)$, implying that this line of thermodynamic instability would be dynamically irrelevant if  dynamical arrest were the only kinetic pathway available to the system.

Performing these calculations at other isochores allows us to determine the two glass transition lines  ($T =  T_\lambda(\phi)$ and $T = T_c(\phi)$) shown in Fig. \ref{Fig2b}, which merge at the bifurcation point $(\phi_b,T_b)$ into a single fluid-glass transition line, and which coincides with that determined by the method of subsection \ref{subsection3.3} (see Fig. \ref{Fig2a}). For completeness, Fig. \ref{Fig3b} illustrates the behavior of the function $\gamma_a(\phi=0.45,T)$, corresponding to an isochore at the right of the bifurcation $(\phi_b,T_b)$. In this case, one finds that $\gamma_a(\phi=0.45,T)$ is infinite for temperatures above the  fluid-glass transition line $T = T_c(\phi=0.45)=0.656$, and  jumps discontinuously to a finite value $\gamma_a(\phi=0.45,T_c^-)\approx 0.025$. To the right of the bifurcation ($\phi>\phi_b$), the function  $\gamma_a(\phi,T)$ is continuous through both, the $\lambda$ line and the spinodal curve, implying that neither of these lines have any dynamical relevance in this regime $\phi>\phi_b$.

\begin{figure*}[ht]
\subfigure{\label{fig:4b}
\includegraphics[scale=0.27]{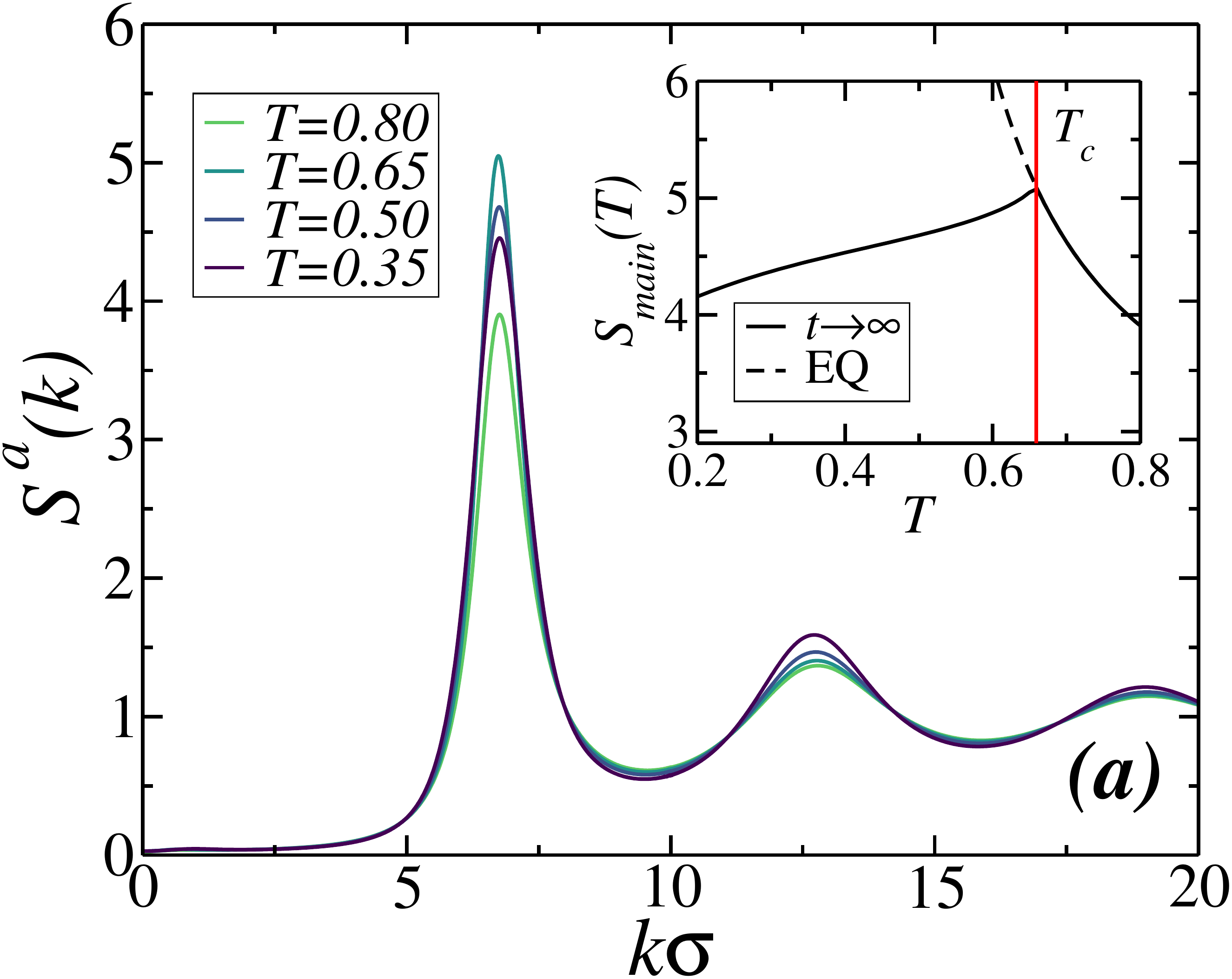}
}
\subfigure{\label{fig:4c}
\includegraphics[scale=0.27]{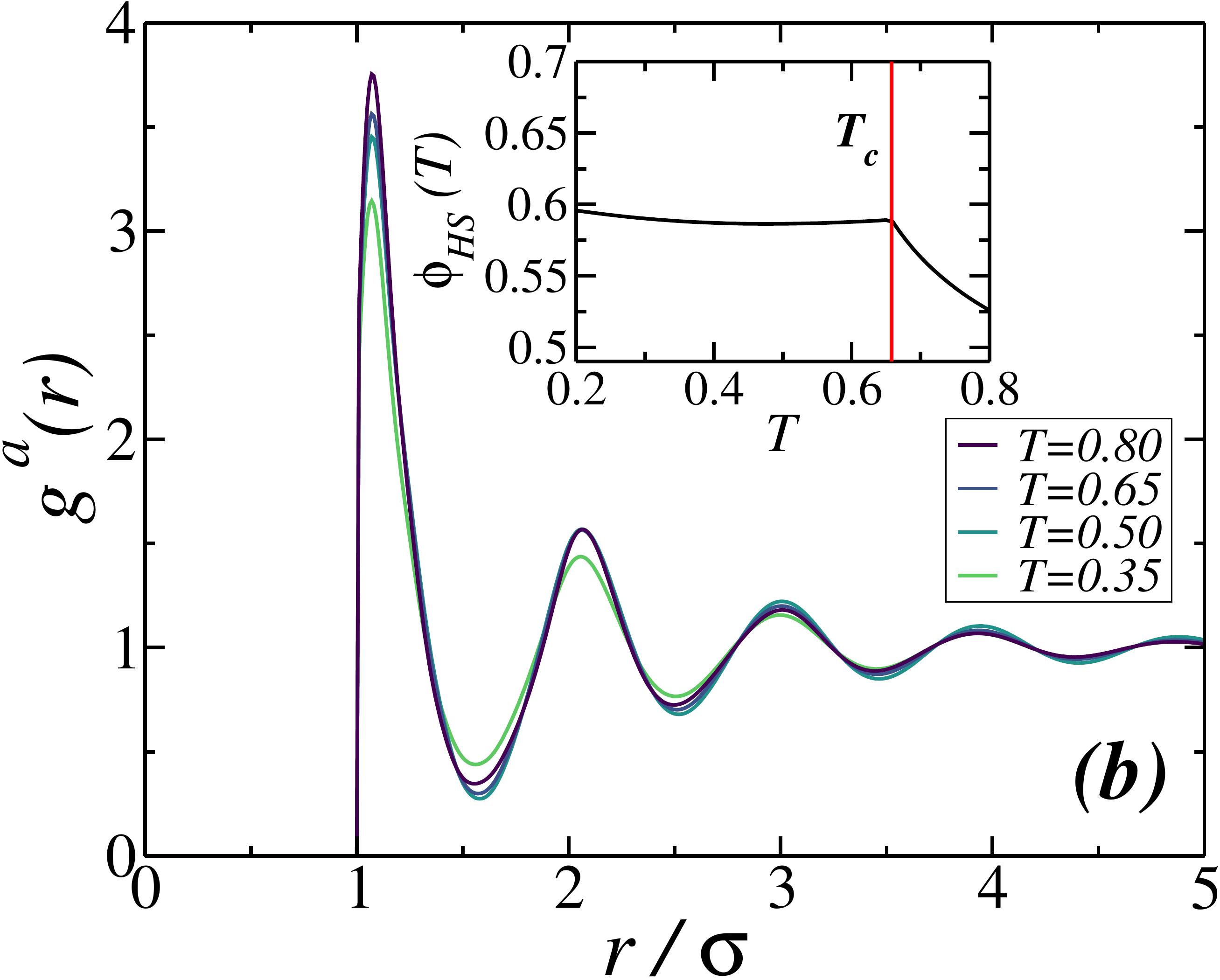}
}
\caption{Behavior of the stationary (a) NESF $S^a(k;\phi,T)$ and (b) NERDF $g^a(r;\phi,T)$, of the HSDY fluid 
($z_1=1$, $z_2=0.5$ and $A=0.5$), at fixed volume fraction $\phi=0.45$ and for different (final) temperatures $T$ above and below the liquid-glass transition $T_c(\phi=0.45)=0.656$, as indicated. The inset of (a) exhibits the $T$ dependence of the height $S^a_{main}(\phi,T) \equiv S^a(k_{main};\phi,T)$ of the main peak of $S^a(k;\phi,T)$ (darkest solid line) and of the equilibrium structure factor $S(k;\phi,T)$ (dashed line). The solid line in the inset of (b) represents the effective hard-sphere volume fractions $\phi_{HS}(\phi=0.45,T)$, determined as described in the text. }
\label{fig4}
\end{figure*}
Let us emphasize that the dynamic arrest scenario summarized by Fig. \ref{Fig2b} appears at first sight to be essentially  identical to that of the same model system in the absence of the repulsive Yukawa interaction (i.e., with $A=0$, see Refs. \cite{olais1,olais2,nescgle8,zepeda}). This is illustrated by comparing Fig. \ref{Fig2b} above, with  Fig. 4 of Ref. \cite{olais1}. There are, however, remarkable fundamental differences. 
The most important of them is the nature of the dynamic arrest transition  line $T = T_\lambda(\phi)$, along with the functions $u_a(\phi,T)$ and $\gamma_a(\phi,T)$, which jump, from infinite values above the $\lambda$ line, to finite values below (type B transition). In the absence of the repulsive term, $A=0$, there is no  $\lambda$ line, and the type B arrest line  $T =  T_\lambda(\phi)$ is replaced by an arrest line $T = T_s(\phi)$ lying along the spinodal curve. This transition happens to be of type A, i.e., the function $\gamma_a (\phi,T)$ passes continuously from infinite values above the line $T = T_s(\phi)$, to finite values below. 

This qualitative difference between these two cases is, of course, associated with a remarkable \emph{structural} difference between the arrested phases predicted immediately below each transition.  Thus,  in the absence of repulsions ($A=0$), the NE-SCGLE theory predicts the formation, below  $T = T_s(\phi)$, of spinodal-decomposition heterogeneities, whose growth is halted only by dynamic arrest mechanisms, as established and discussed in detail in Refs.  \cite{olais1,olais2}.  In contrast, in the presence of repulsions ($A>0$), below $T =  T_\lambda(\phi)$ the NE-SCGLE theory predicts the dynamic arrest due to the repulsion between finite-size clusters, 
as we now establish by analyzing the structural properties predicted by the NE-SCGLE theory.  

\begin{figure*}[ht]
\subfigure{\label{fig:5a}
\includegraphics[scale=0.27]{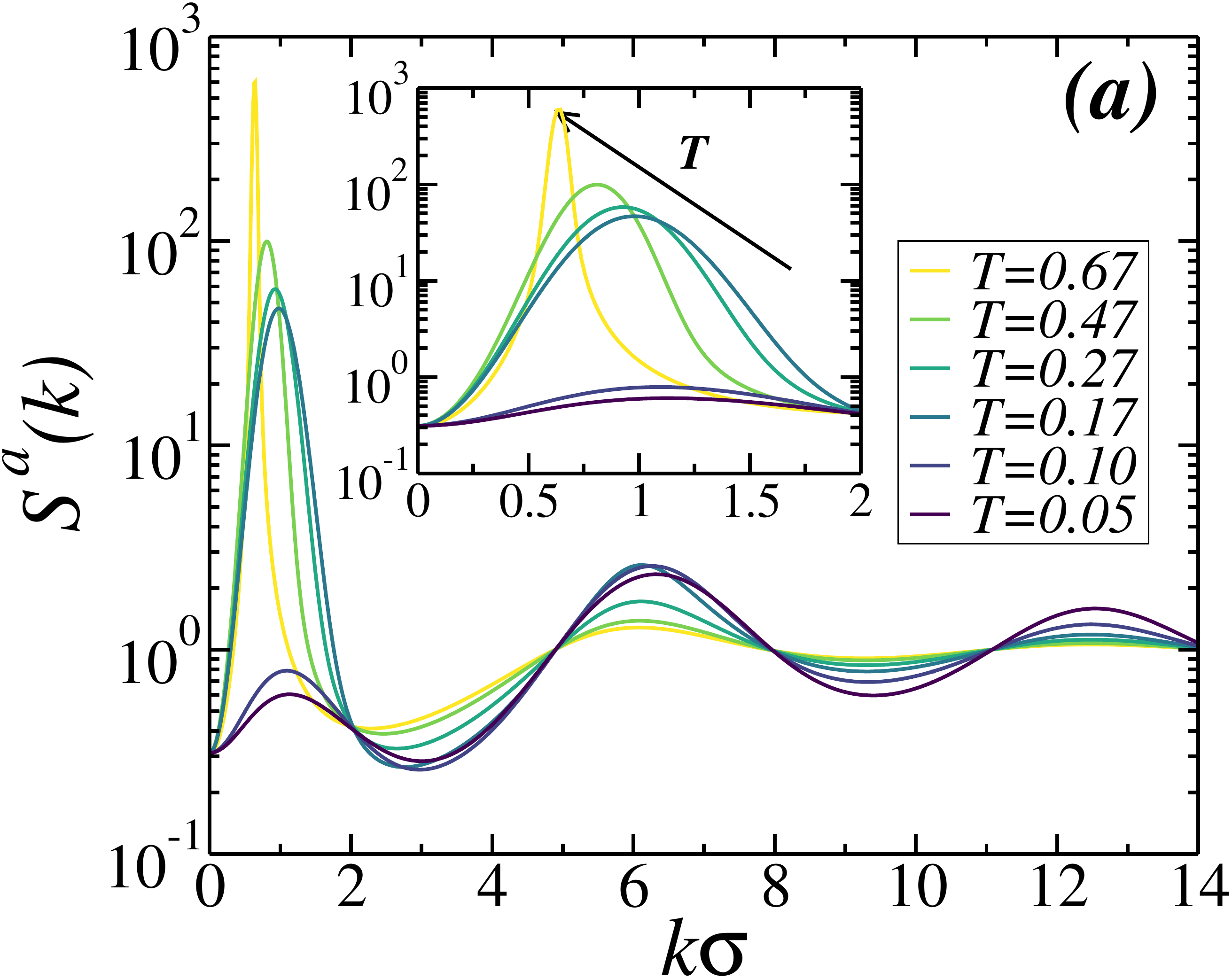}
}
\subfigure{\label{fig:5b}
\includegraphics[scale=0.27]{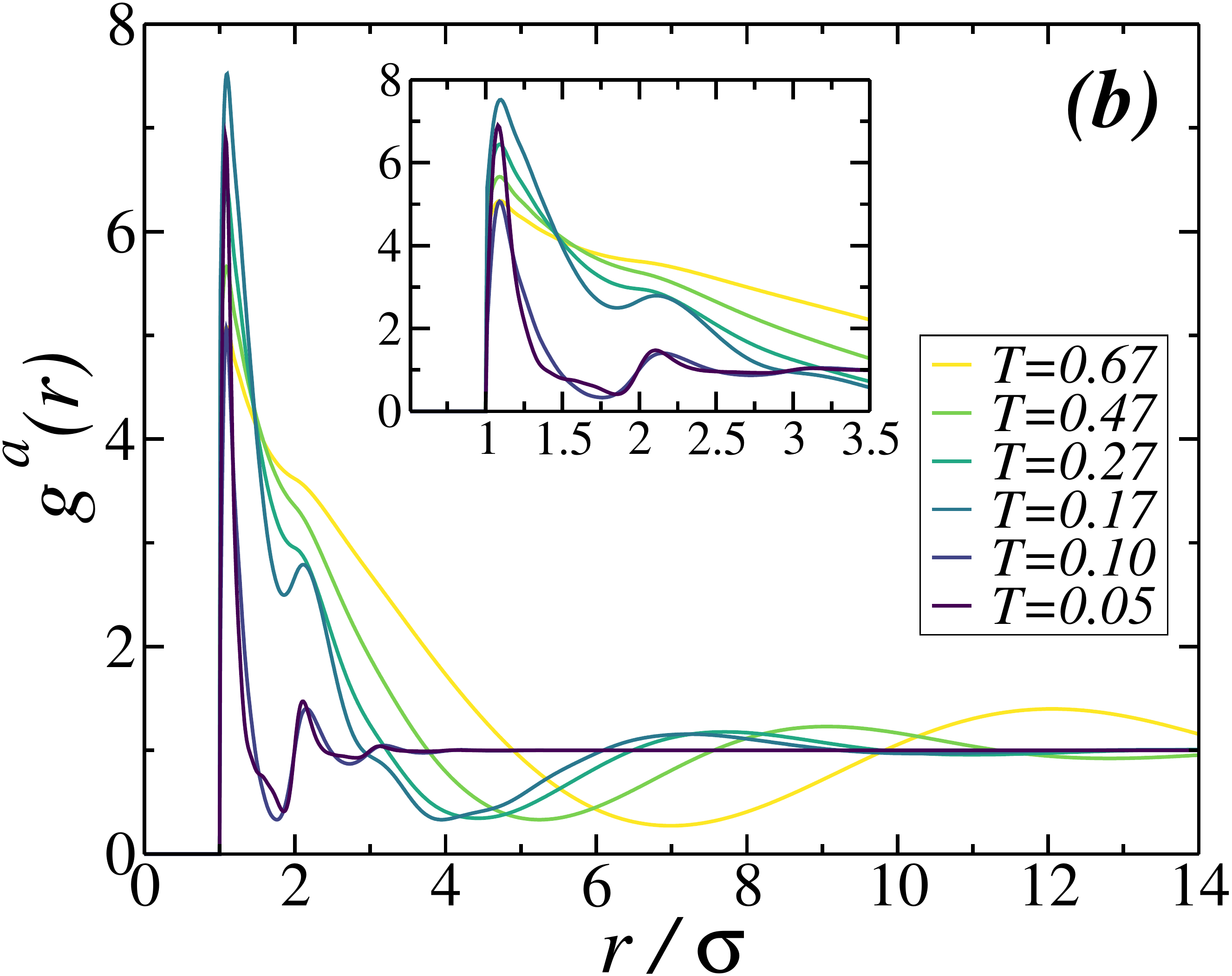}
}
\caption{Behavior of the stationary (a) NESF $S^a(k;\phi,T)$ and (b) NERDF $g^a(r;\phi,T)$, of the HSDY fluid ($z_1=1$, $z_2=0.5$ and $A=0.5$), at fixed volume fraction $\phi=0.15$ and for different temperature quenches below $T_{\lambda}(\phi=0.15)=0.6733$. The insets are auxiliary zooms to emphasize both the amplification of structural correlations around $k_{\lambda}$ (or $r_{\lambda}$, according to the case) , and the shift of $k_{\lambda}$ ($r_{\lambda}$) towards smaller (larger) values with decreasing $T$. In both figures, one notices two qualitatively different behaviors: for shallow quenches, strong amplifications at low wave vector $k_{\lambda}$ (or $r_{\lambda}$), and for deeper quenches, only a moderate amplification at $k\approx 2\pi/\sigma$. }
\label{fig5}
\end{figure*}

\section{Stationary structure factor $S^a(k)$.}\label{section4}

Let us now discuss the structural properties of the HSDY model predicted by the NE-SCGLE theory, as described by the solution of Eq. (\ref{dsktenst}) for $S(k;t)$. As carefully discussed in Ref. \cite{olais1},  the time-dependent NESF $S(k;t)$ is given by  $S(k;t)=S^*(u(t))$, where the function $S^*(u)$ is defined in Eq. (\ref{solsigmadku}), and with the variable $u(t)$ being the ``material'' time $u(t)\equiv \int_0^t b(t')dt'$ \cite{peredo1}. Before discussing the structural aging described by the full $t$-dependence of $S(k;t)$, however, it is useful to consider its stationary, long-time asymptotic limit $S^a(k)\equiv \displaystyle{\lim_{t\to\infty}} S(k;t)$,  given by \cite{olais1}
\begin{equation}
\begin{split}
S^a(k;\phi,T) \equiv
[n\mathcal{E}(k;\phi,T)]^{-1}+\left \{ S_i(k)-
[n\mathcal{E}(k;\phi,T)]^{-1}\right \}\times \\
e^{-2k^2D_0
n\mathcal{E}(k;\phi,T)u_a(\phi,T)}, 
\end{split}
\label{solsigmadkua}
\end{equation}
with the initial condition $S_i(k)$ chosen here as the SF at infinite $T$, i.e, $S_i(k)=S(k;\phi,T=\infty)=S^{HS}(\phi)$. 

Clearly, when the system is quenched to a state point $(\phi,T)$ in the ergodic region (infinite $u_a(\phi,T)$ and $\gamma_a(\phi,T)$), the system will equilibrate, and $S(k;t)$ will asymptotically attain its equilibrium value $S^a(k;\phi,T)=S(k;\phi,T)$. The salient features of $S(k;\phi,T)$ have been discussed thoroughly in the literature \cite{seargelbart,LiuChenChenJCP05, archerpinievansreatto07}, and were briefly illustrated in Fig. \ref{fig1}. The case of interest that the NE-SCGLE theory now allows us to discuss is, of course, when the state point $(\phi,T)$ lies in one of the non-ergodic regions (finite $u_a(\phi,T)$ and $\gamma_a(\phi,T)$), in which the system will no longer be able to equilibrate. Instead, $S(k;t)$ is expected to attain the non-equilibrium value $S^a(k;\phi,T)$ given by Eq. (\ref{solsigmadkua}).

Figs. \ref{fig4} and \ref{fig5}  illustrate the dependence on the depth of the quench of both, the NESF $S^a(k;\phi,T)$ and the corresponding NERDF $g^a(r;\phi,T) \equiv 1+ h^a(r;\phi,T)$, with the NETCF $h^a(r;\phi,T)$ being the inverse FT of  $[S^a(k;\phi,T)-1]/n$. To  discuss the high-density--high-temperature scenario, Figs. \ref{fig4}(a) and  \ref{fig4}(b) exhibit the behavior predicted for
these structural properties  along the isochore $\phi=0.45$, which lies well to the right of the bifurcation point $(\phi_b,T_b)$, as a function of $T$. In Fig. \ref{fig4}(a) we notice first the non-monotonic $T$-dependence of the hard-sphere--like correlations, represented by the height $S^a_{main}(\phi,T) \equiv S^a(k_{main};\phi,T)$ of the main peak of $S^a(k;\phi,T)$, located at $k=k_{main} \approx 2\pi/\sigma$. As summarized in the inset of this figure, this property (described there by the solid line) first increases when $T$ is decreased still in the ergodic regime,  $T>T_c(\phi)$, but  as $T$ crosses the liquid-glass line, it decreases with smaller $T$. Note that the SF $S(k;\phi,T)$ still exists for $T\le T_c(\phi)$ and the height $S(k_{main};\phi,T)$ of its main peak continues to increase as $T$ decreases below $T_c(\phi)$ (dashed line). The NE-SCGLE theory, however,  predicts that below $T_c(\phi)$,  this equilibrium $S(k;\phi,T)$ will be unreachable in practice due to kinetic considerations \cite{nescgle3}. 

Let us notice also the qualitative resemblance of the non-equilibrium structure represented by $S^a(k;\phi,T)$  with the equilibrium structure of a fluid of particles with only excluded volume interactions.  This is also observed in the dependence of the NERDF  $g^a(r;\phi,T)$ illustrated in  Fig. \ref{fig4}(b). To quantify this resemblance, let us borrow the notion of structural equivalence between equilibrium soft and hard-sphere fluids proposed in Ref. \cite{dynamicequivalence,prl,pre}, and let us apply it to establish the structural equivalence between our non-equilibrium HSDY system and the equilibrium HS liquid. For this, we define an effective HS fluid whose RDF $g_{HS}(r;\phi_{HS})$ matches $g^a(r;\phi,T)$ at least in the height $g^a_2(\phi,T)\equiv g^a(r_2;\phi,T)$ of its second maximum, located at $r_2 \approx 2\sigma$.


This procedure assigns an effective HS volume fraction  $\phi_{HS}$ to each state point $(\phi,T)$. In the inset  of Fig. \ref{fig4}(b) we plot the function $\phi_{HS}=\phi_{HS}(\phi,T)$ for $\phi=0.45$, which reveals  that in the ergodic region $\phi_{HS}(\phi,T)$ increases with the depth of the quench until reaching a maximum similar in magnitude to the well-known hard-sphere glass transition volume fraction $\phi^{HS}_c=0.58$ at $T\approx T_c$. Below the liquid-glass transition line $T =T_c(\phi)$, however, we observe that $\phi_{HS}(\phi,T)$ remains relatively constant, reminding the $T$-dependence of the inverse square localization length $\gamma_a^{*-1}(\phi,T)$ below $T_c$ (see Fig. \ref{Fig3}(b)). These results provide non-equilibrium evidences to conclude that, when the fluid-glass transition is crossed along isochores in the high-density--high-temperature regime, to the right of the bifurcation point $(\phi_b,T_b)$, and far from the $\lambda$ line, dynamical arrest is mainly driven by the excluded volume interactions and caging mechanisms.  As a side observation, let us recall that the definition of the function $\phi_{HS}=\phi_{HS}(\phi,T)$ was based on the mapping of the non-equilibrium structure of the HSDY system onto the purely HS equilibrium structure. Such mapping is characteristic of the so-called ``HS dynamic universality class'' \cite{prl,pre}. This leaves open the provocative possibility that, within adequate scalings, the asymptotic dynamics in this regime could also be mapped onto the equilibrium dynamics of the HS system.

We now move to the complementary (low-density--low-temperature) regime, and consider the same isochore as in Fig. \ref{fig1}, $\phi=0.15$, which lies well to the left of the bifurcation point. Going back to that figure, we recall that, along this low-density isochore, the salient feature, amply 
discussed in the literature  \cite{seargelbart,LiuChenChenJCP05,archerpinievansreatto07,ruizzaccarelli}, was the emergence of the low-$k$ (``cluster'') peak of the SF $S(k;\phi,T)$, centered at the wave vector denoted as $k_\lambda$.
There, we learned that the height $S(k_\lambda;\phi,T)$ of this low-$k$ peak  increases monotonically as the temperature is decreased until reaching the $\lambda$ line, where it diverges. Fig.  \ref{fig5}(a) now complements this equilibrium story with its non-equilibrium continuation below the $\lambda$ line, where the SF $S(k;\phi,T)$ ceases to exist, but the second, NESF $S^a(k;\phi,T)$ given by Eq. (\ref{solsigmadkua}), emerges. This new non-equilibrium stationary solution represents, of course, the formation of amorphous dynamically arrested materials. 

Fig.  \ref{fig5}(a) illustrates how $S^a(k;\phi,T)$ changes as the final temperature is lowered, now below $T_\lambda (\phi)$. The main trend observed is highlighted in the inset, and refers to the notorious decrease with the quench depth, of the height $S^a_\lambda(\phi,T)\equiv S^a(k_\lambda;\phi,T)$ of the low-$k$ peak of $S^a(k;\phi,T)$, starting from its maximum value attained at $T\approx T_{\lambda}$ (e.g., $S^a_\lambda(\phi=0.15,T=0.67)\approx 500$), down to a value of $\approx 45$ for $T$ approaching the glass-glass transition $T_c=0.165$ from above. Decreasing $T$  below $T_c$, we observe that the height $S^a_\lambda(\phi=0.15,T)$ of the small-$k$ peak drops precipitously, down to values smaller than the height of the main (``particle-particle'') peak at $k\approx 2\pi /\sigma$. While this happens to the height $S^a_\lambda(\phi,T)$, we also observe a shift of the position $k_\lambda$  of this peak, to larger wave-vectors, as $T$ decreases below $T_\lambda$. This indicates that the size $\xi=2\pi/k_\lambda$ of the corresponding spatial heterogeneities are largest at $T\approx T_{\lambda}$ (e.g., $\xi \approx 11\sigma$ at $T=0.67$, slightly below $T_\lambda$), and decreases with the depth of the quench, down to approximately  $6\sigma$  at  $T=0.17$ (slightly above $T_c$).

\begin{figure*}[ht]
\subfigure{\label{fig:6a}
\includegraphics[scale=0.2]{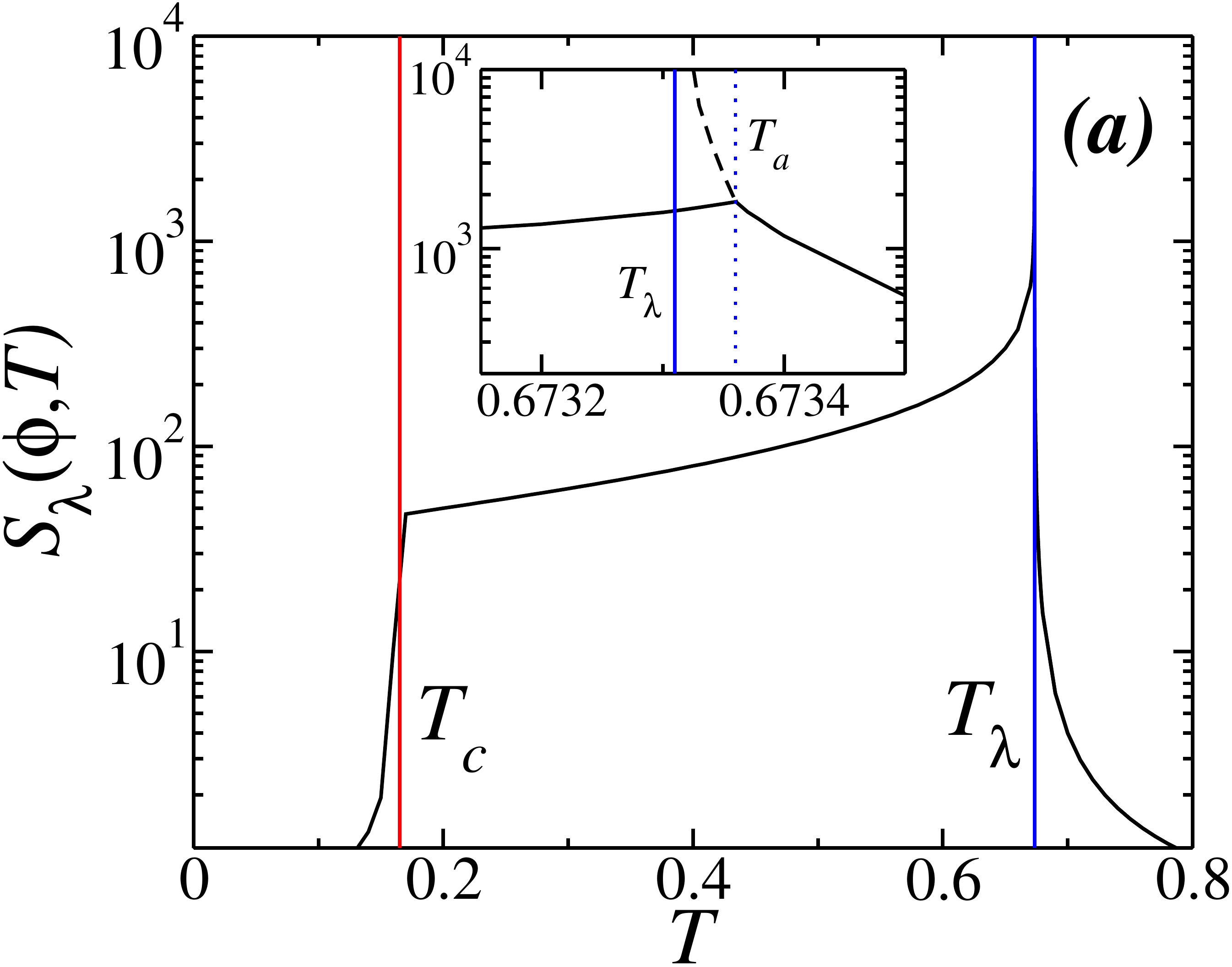}
}
\subfigure{\label{fig:6b}
\includegraphics[scale=0.2]{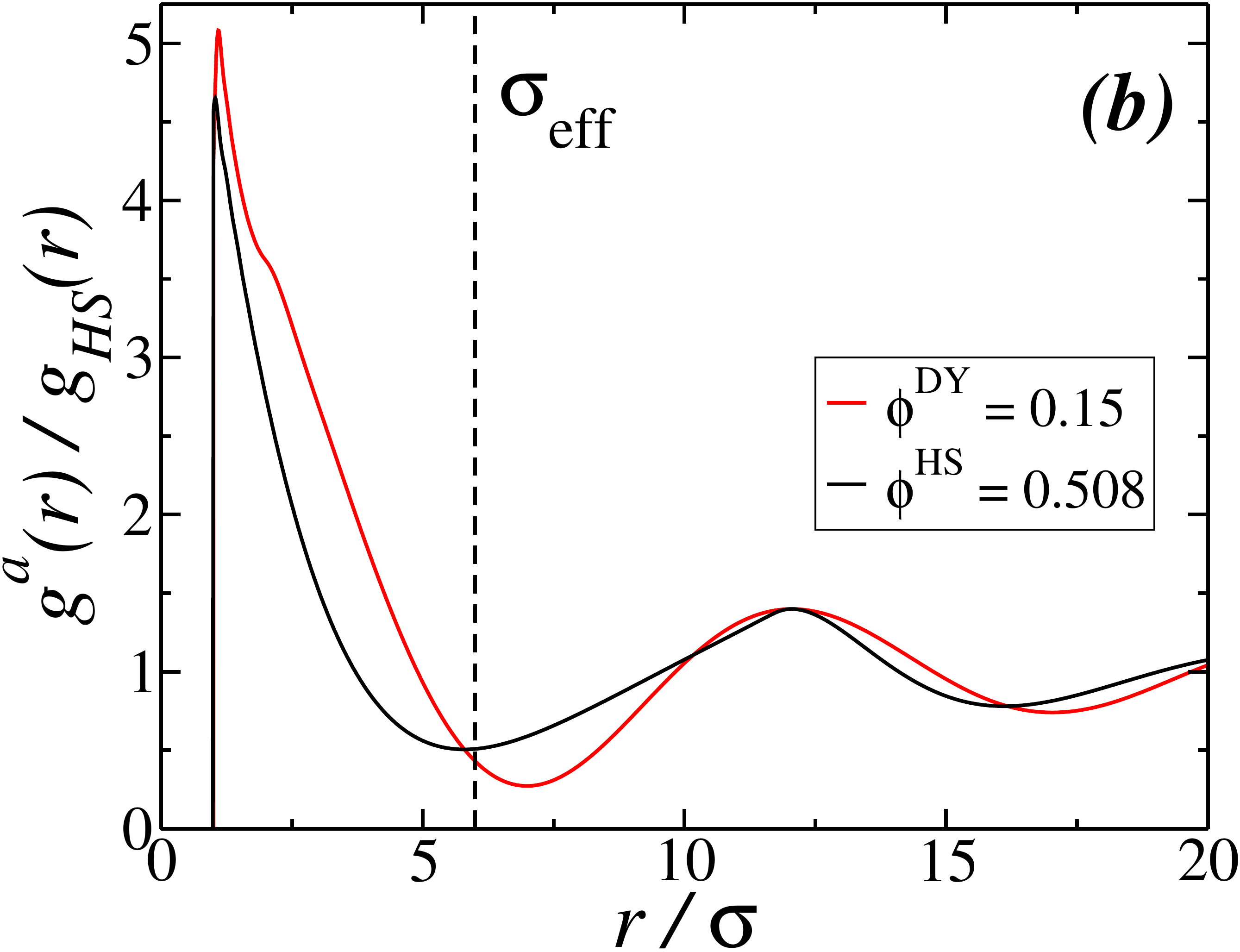}
}
\subfigure{\label{fig:6c}
\includegraphics[scale=0.2]{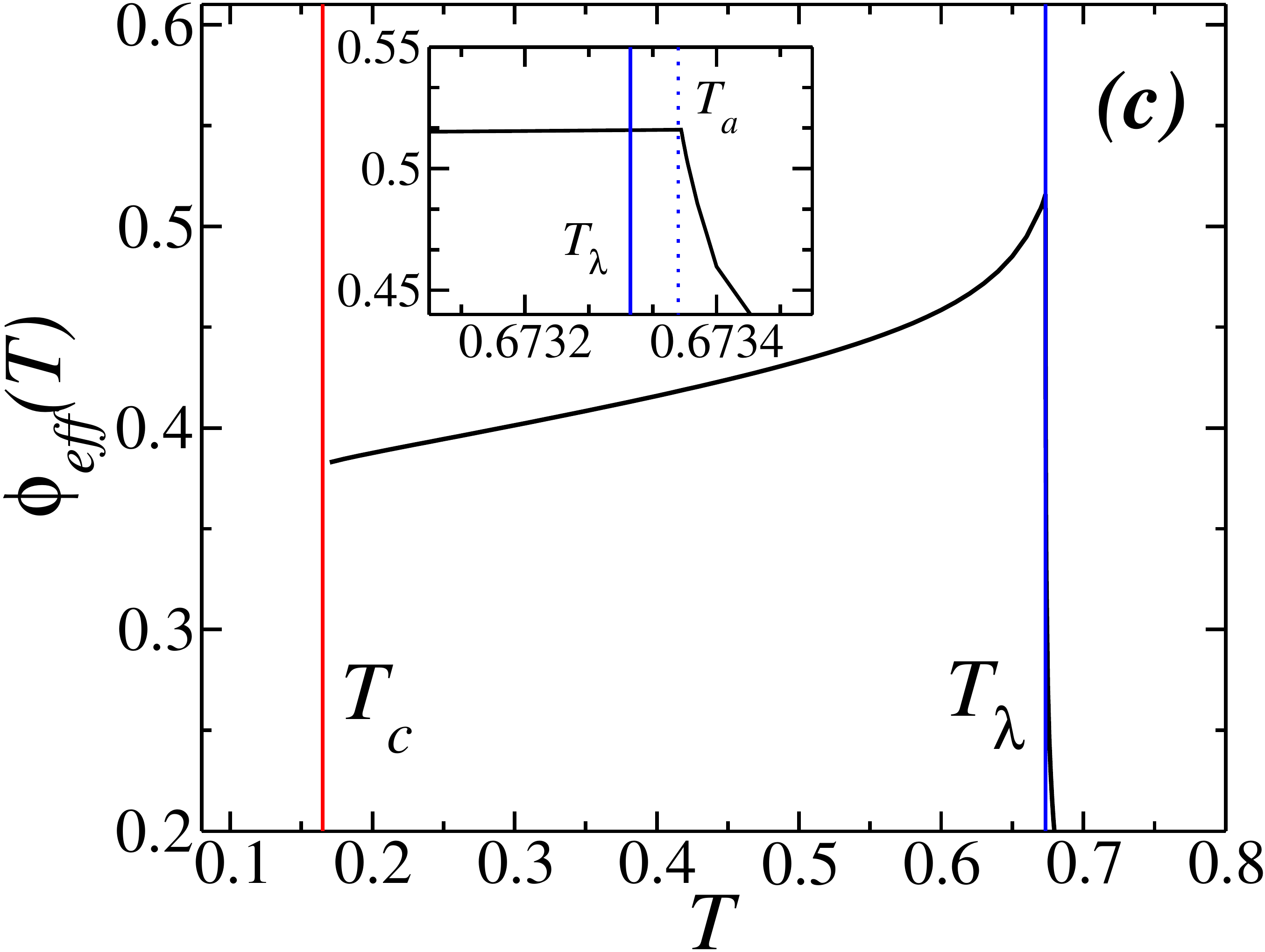}
}
\caption{(a) $T$-dependence of the height $S^a_{\lambda}(\phi,T)\equiv S^a(k=k_{\lambda};\phi,T)$ of the NESF $S^a(k:\phi,T)$ along the isochore $\phi=0.15$, covering temperatures from below the glass transition point $T_c$ up to temperatures above the fluid-glass-transition temperature $T_{a}=0.67336>T_{\lambda}=0.67331$ (see the text). The inset considers an amplification around $T_a$, used to emphasize the deviations of $S^a_{\lambda}(\phi,T)$ (solid line) from the equilibrium value $S(k_{\lambda};\phi;T)$ (dashed line) which diverges at $T_{\lambda}$. (b) Comparison of the behavior of the asymptotic value of the NERDF $g^a(r;\phi=0.15,T=0.67)$ (red solid line) and the effective hard-sphere RDF $g(r;\phi_{HS})$ (black solid line) used to match the height of the second maximum (see the text). (c) T- dependence of the volume fraction $\phi_{eff}$ of the effective hard-sphere fluid used to match the RDF for temperatures below the $T_a$ transition line and slightly above the $T_c$ transition (see the text).
}
\label{fig6}
\end{figure*}

The same information is presented in Fig.  \ref{fig5}(b), but in terms of the stationary NERDF $g^a(r;\phi,T)$. There we see that  the long-range minimum in the RDF $g(r;\phi,T)$ located at a distance $r_{min}\approx 7\sigma$, recalling the lower temperature in Fig. \ref{fig1}(b)  ($T=0.68 > T_\lambda$), now continues to appear in the NERDF $g^a(r;\phi,T)$ for quenches below $T_\lambda$. This is illustrated by the shallowest quench of Fig.  \ref{fig5}(b), with final temperature $T=0.67$,  whose NERDF $g^a(r;\phi,T)$ shows a much more pronounced long-range minimum at $r_{min} \approx 7\sigma$, and a high long-range second maximum $g_2^a(\phi,T)\equiv g^a(r_{\lambda};\phi,T)$ at $r_{\lambda}\approx 12\sigma$. These long-ranged minimum and maximum describe the cluster layering around a central particle (itself in a cluster). One notices, in addition, that the relevance of these features, measured by the heights $S^a(k_\lambda ;\phi,T)$ or $g_2^a(r_{\lambda};\phi,T)$, is largest in the neighborhood of the $\lambda$ line and decreases with the quench depth.

To have a more comprehensive summary of these trends in the behavior of $S^a(k;\phi,T)$, in Fig. \ref{fig6}(a) we plot $S^a_{\lambda}(\phi=0.15,T)$ as a function of $T$, covering  temperatures from below the glass-glass transition temperature $T_c$ up to temperatures above $T_\lambda$. There we see that it is mostly in the temperature interval $T_c< T < T_\lambda$ where  $S^a_{\lambda}(\phi,T)$ attains very large values, decreasing  catastrophically outside this interval. Regarding the decay of $S^a_{\lambda}(\phi,T)$ with $T$ in the close neighborhood of $T_\lambda$, if the dynamic arrest transition occurred exactly at $T_\lambda$, we should have a mathematical  discontinuity right at  $T_\lambda$, since $S^a_{\lambda}(\phi,T)$ remains finite as $T$ approaches $T_\lambda$ from below, but diverges when $T$ approaches  $T_\lambda$ from above.

A close inspection of such a singular prediction, however, revealed a rather unexpected scenario, illustrated in the inset of Fig. \ref{fig6}(a): as it turns out, the dynamic arrest transition does not occur at $T=T_\lambda$, but at a different, slightly larger temperature, which we denote $T_a$. More precisely, $T_\lambda= 0.67331$, whereas $T_a=0.67336$ (clearly,  to detect this difference we needed a very high numerical resolution). However, the consequence is that the dynamic arrest transition preempts the occurrence of the divergence of $S^a_{\lambda}(\phi,T)$, thus explaining the practical impossibility of its observation.  Hence, although quantitatively modest, this observation may have far-reaching fundamental implications regarding the theoretical understanding of the accurate descriptions of cluster formation and dynamic arrest in SALR systems. In fact, although scanning the state space spanned by the potential parameters ($z_1$, $z_2$, and $A$) is out of the scope of the present work, let us mention that preliminary extensions of the present NE-SCGLE calculations reveal that the minute difference between $T_\lambda$ and $T_a$, becomes more significant when the range of the attraction decreases. We thus expect that a more detailed study in this direction will allow a more direct contact with the phenomenology of cluster formation revealed by simulations \cite{sciortinomossa,charbonneaureichmanPRE07}, performed for much narrower attractive wells than that discussed here. In this paper, however, we shall continue neglecting this quantitative difference between $T_a$ and  $T_\lambda$, and will continue describing the dynamic arrest transition as occurring at the temperature $T_\lambda$.

On a different subject, let us now refer to the early and sharp analysis by  Sciortino et al. \cite{sciortinomossa}, who modeled the effective long-range cluster-cluster interaction as the superposition of the particle-particle Yukawa repulsion between the particles that form two interacting clusters. This led to a renormalized cluster-cluster  Yukawa repulsion, capable of inducing dynamic arrest into a Wigner glass of clusters, very much as the experimentally observed \cite{82LindsayChaikinJCP,89SirotaPRL} Wigner glasses formed by  individual charged colloidal particles. Along this line of thought, the long-range minimum and maximum of $g^a(r;\phi,T)$, in Fig. \ref{fig5}(b) should correspond to the first minimum and maximum of the RDF of a repulsive Yukawa fluid. However, since the repulsive Yukawa fluid is structurally equivalent to the hard-sphere fluid \cite{prl,pre}, we can also pursue this idea using the structural equivalence of the fluid of clusters with the hard sphere liquid.


This is best explained in Fig. \ref{fig6}(b), which compares the NERDF $g^a(r;\phi,T)$ (red solid line) corresponding to the shallowest quench in Fig. \ref{fig5}(b) (final temperature $T=0.67$, slightly below $T_\lambda=0.6733$), with the RDF $g(r;\phi_{HS})$ (black solid line) of an effective hard-sphere system that matches the height   $g^a_2(\phi,T)\equiv g^a(r_2;\phi,T)$  of the long-range second maximum at $r_2$ of $g^a(r;\phi,T)$. This comparison determines the HS volume fraction $\phi_{HS}(\phi,T)$ of an effective HS system structurally equivalent to the cluster-cluster correlations in the RDF $g^a(r;\phi,T)$ of the SALR system at the final state point $(\phi,T)$. Fig. \ref{fig6}(c) exhibits the value $\phi_{eff}\equiv \phi_{HS}(\phi=0.15,T)$ thus computed, as a function of the final temperature $T$. 

As observed in Fig. \ref{fig6}(c),   $\phi_{eff}$  can be determined even in the equilibrium regime ($T>T_\lambda$), but only in the immediate neighborhood of $T_\lambda$, where it has a sharp increase as $T$ approaches $T_\lambda$ from above, and  where it reaches its maximum value (slightly larger than 0.5). For $T<T_\lambda$, we have that $\phi_{eff}$ decreases with decreasing $T$, down to a value $\phi_{eff}\lesssim 0.45$ in the neighborhood of the glass-glass transition temperature $T_c$. Thus,  in the interval $T_c < T < T_\lambda$, the quantitative value of $\phi_{eff}$ in Fig. \ref{fig6}(c) is clearly smaller than the HS glass transition volume fraction $\phi_c\approx 0.58$, and hence, cannot account for the dynamic arrest condition. This means that the cluster-cluster caging mechanism contributes to the arrest, but it is not sufficient. Hence, other mechanisms must come into play in this regime. 

Those additional mechanisms are also suggested by the results for $g^a(r;\phi,T)$ in Fig. \ref{fig5}(b). There we see that, as $T$ decreases sufficiently, the true hard-sphere interactions manifest themselves in the emergence of a local maximum in $g^a(r;\phi,T)$ at  $r\approx 2$, indicating growing particle-particle (not cluster-cluster) hard-sphere correlations. As the results in Fig. \ref{fig5}(b) indicate, these become the dominant correlations for $T$ below the glass-glass transition temperature $T_c$.

In summary, we can say that, in contrast with the high-density regime illustrated in Fig. \ref{fig4}, whose dynamic arrest was strongly dominated by the excluded volume forces, in the low-density--low-temperature regime, illustrated by the isochore $\phi=0.15$, the dynamic arrest  is the result of a much subtler and complex interplay between the three components of the interaction: the excluded volume, the long-range repulsion and the shorter-ranged attraction.

\section{Structural aging: $t$-dependence of $S(k;t)$ and $g(r;t)$.}\label{section5}

At this point, it is important to emphasize that the equilibrium structure 
factor and phase diagram of Fig. \ref{fig1}, as well as  the glass transition diagrams of Fig. \ref{Fig2} and the structural properties $S^a(k;\phi,T)$ and $g^a(r;\phi,T)$ just discussed, only refer to the infinitely long time asymptotic limit, whose experimental observability is only possible, in practice, if we wait longer than the longest relaxation times  of the system. This may be almost trivial when the system is able to relax to equilibrium, which obviously makes the asymptotic equilibrium  scenario verifiable within practical waiting times (and hence, much more familiar and representative of ordinary experience). Under conditions of dynamical arrest, however, the longest relaxation times may actually be too long (ideally, infinite), and the system might then remain in non-equilibrium conditions within any practical observation time. This, in turn, may render the asymptotic glass transition scenario impossible to verify in practice. 

This practical impossibility could impede us to appreciate the value of these asymptotic predictions. Such impediment, however,  can be removed by considering the full $t$-dependent solutions of the NE-SCGLE equations (\ref{dsktenst})-(\ref{fluctsquench}), which predict what one would measure at the finite and practical waiting times $t$ involved in any realistic experiment or simulation. From this kinetic perspective, the fundamental role of the ideal asymptotic scenario may be best appreciated, as we now illustrate with the aging of the structural properties. Of course, the full $t$-dependent solution also yields a wealth of information regarding, for example, the dynamic and rheological properties, thus providing a more powerful resource to understand the nature of the glassy phases described by the asymptotic glass transition scenario.

\begin{figure*}[ht]
\subfigure{\label{Fig7a}
\includegraphics[scale=0.2]{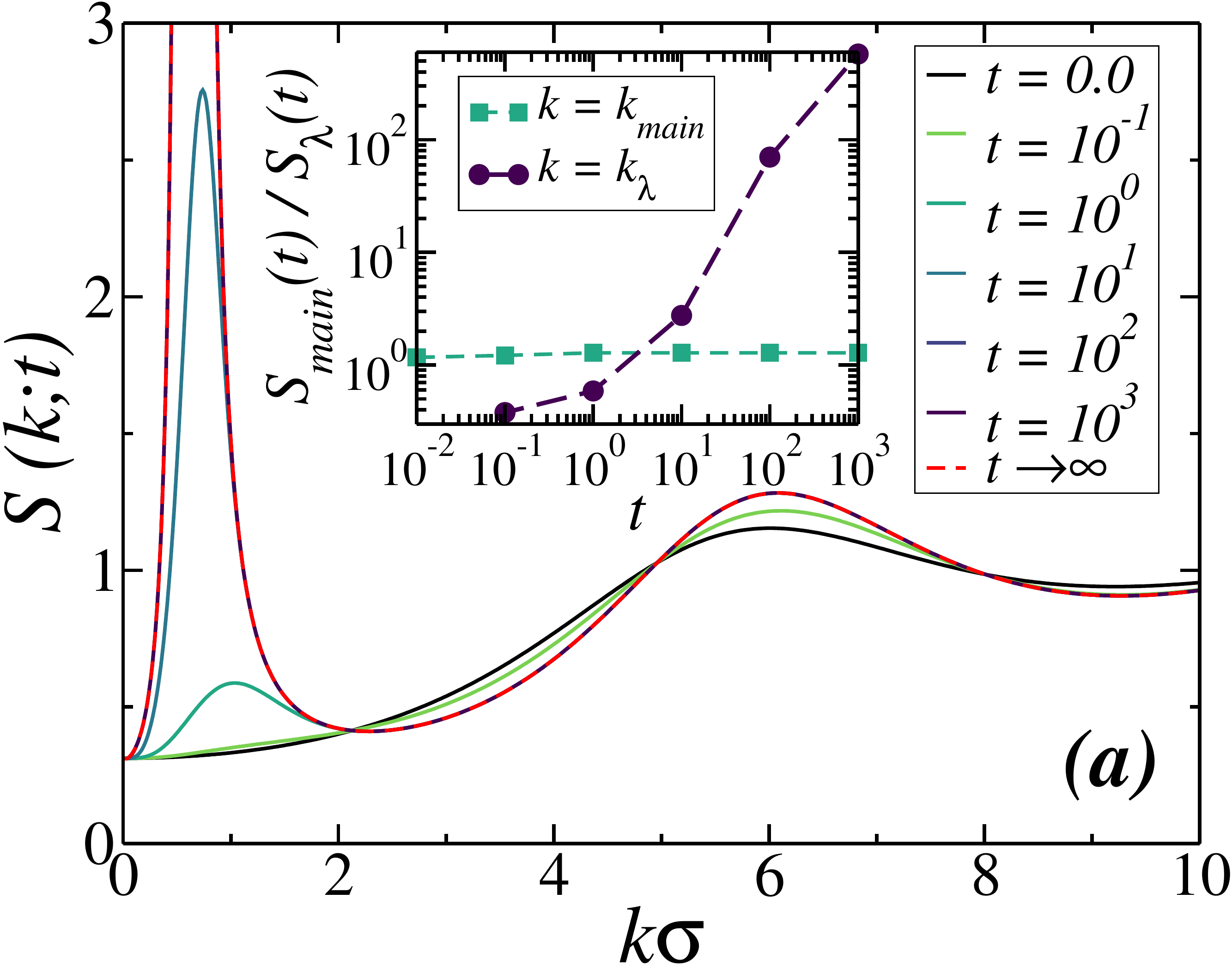}
}
\hspace{10pt}
\subfigure{\label{Fig7b}
\includegraphics[scale=0.2]{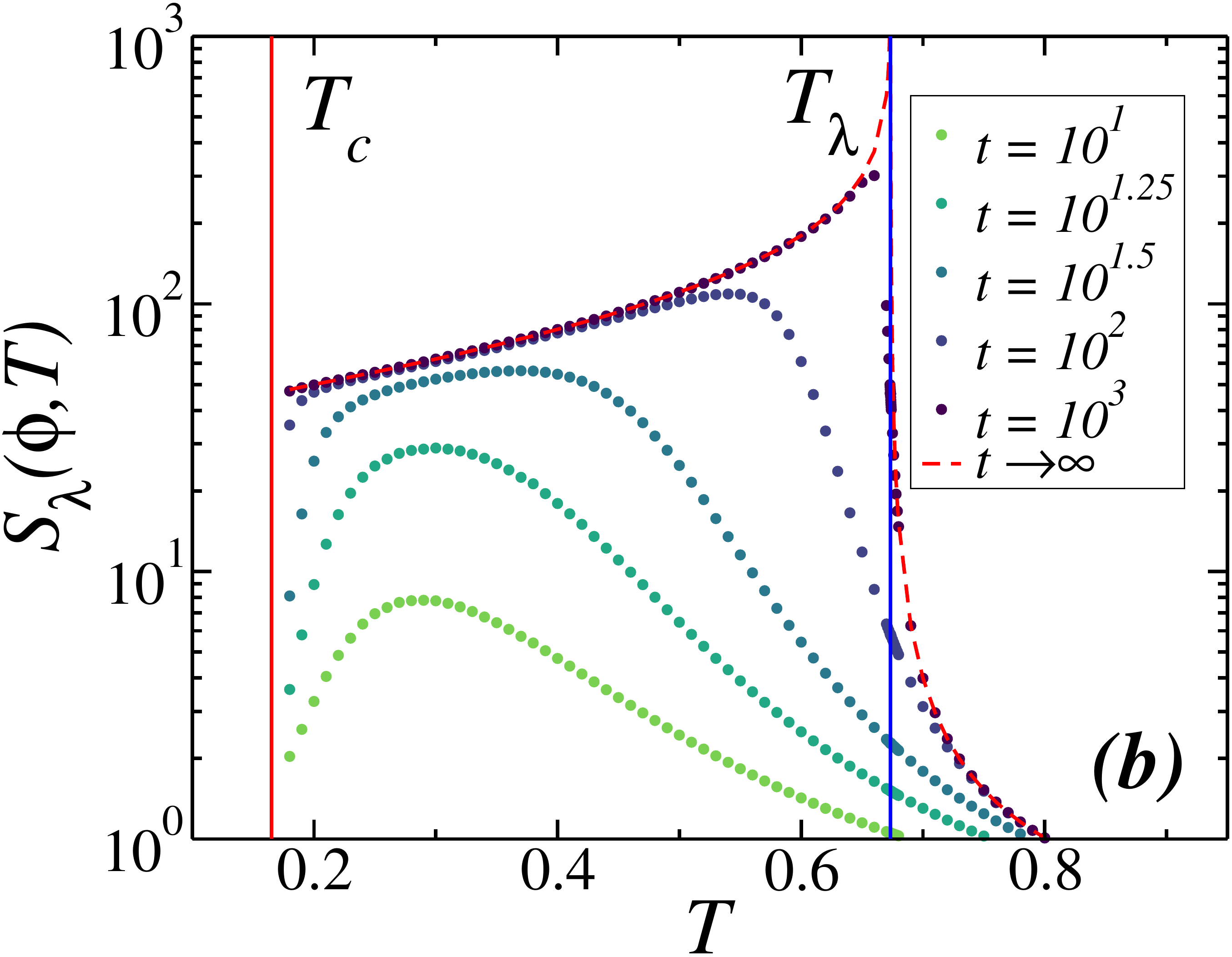}
}
\subfigure{\label{Fig7c}
\includegraphics[scale=0.2]{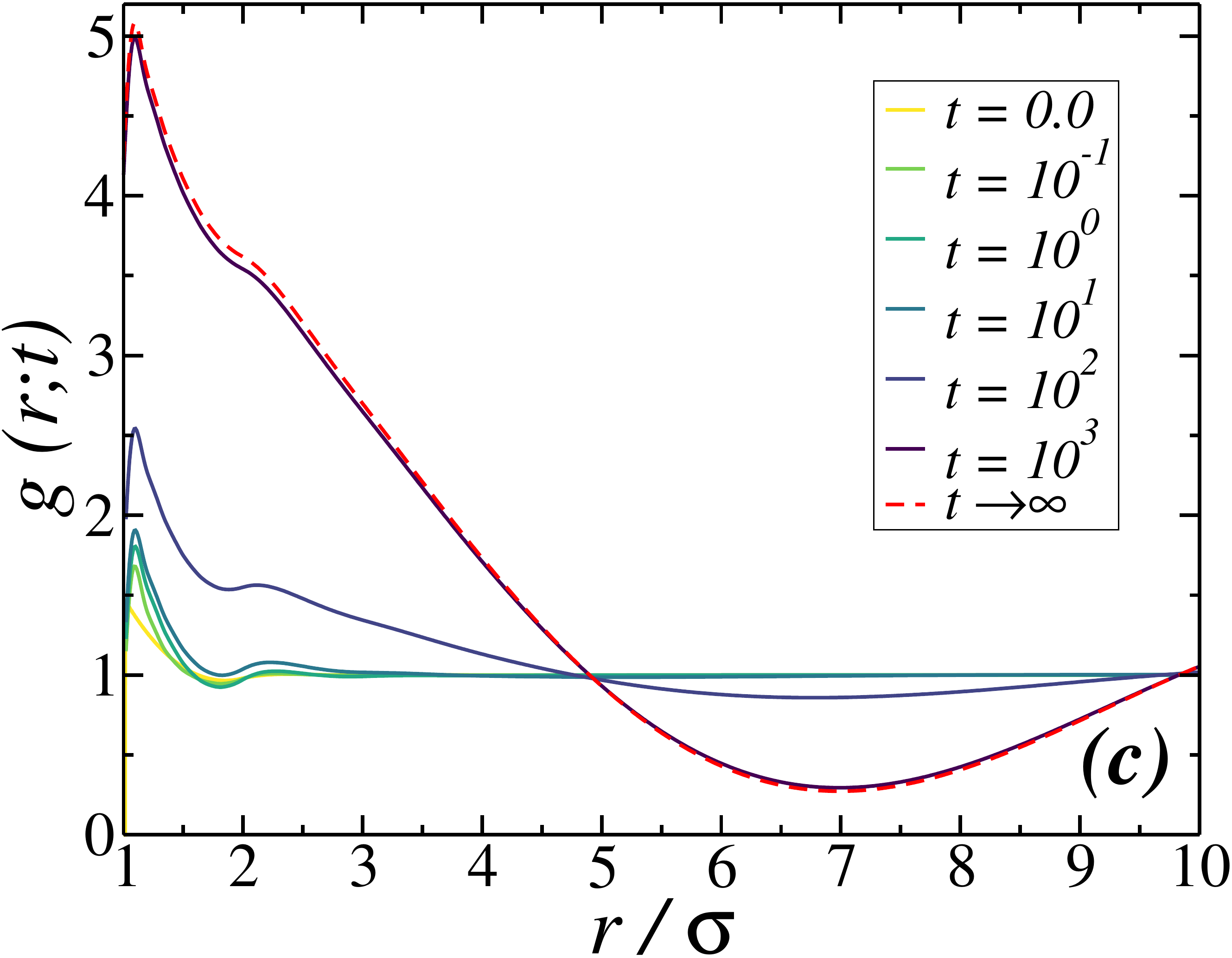}
}
\caption{(a) Sequence of snapshots describing the $t$-evolution 
of the NESF $S(k;t)$ for a quench of the HSDY system, at $\phi=0.15$, from the same initial value as before, and towards 
the final temperature $T=0.67$ (right below $T_{\lambda}(\phi=0.15)$). The inset compares the time evolution 
of the two peaks $S_{main}(t)$ (solid squares) and $S_{\lambda}(t)$
(solid circles). (b) T-dependence of $S_{\lambda}$ for a sequence of 
values of waiting times $t$, as indicated, along with the asymptotic
long time value $S_{\lambda}(t\to\infty)$, also shown in Fig. \ref{fig6}(a). (c) Corresponding sequence of snapshots for the NERDF $g(r;t)(\phi,T)$, along with the asymptotic value $g(r)^a(\phi,T)$.}
\label{fig7}
\end{figure*}

To illustrate the structural aging predicted for the HSDY system, Fig. \ref{fig7}(a) plots the snapshots of the NESF $S(k;t)$ as a function of $k$, for a sequence of representative values of the waiting time $t$, after its instantaneous quench to the final temperature $T=0.67$  along the isochore $\phi=0.15$. These results correspond to the shallowest quench illustrated in Fig. \ref{fig5}, and exhibit the gradual enhancement of the correlations, starting from the chosen initial state  $S(k;t=0;\phi,T)=S^{eq}(k;\phi,T=\infty)=S^{HS}(k;\phi)$ (black solid line), and ending at the asymptotic solution $S^a(k;\phi,T)=S(k;t=\infty;\phi,T)$ (red dashed line). They also illustrate the kinetic build-up of a rather fast, but very modest, increase around the main peak at $k=k_{main} \approx 2 \pi/\sigma$ of $S(k;t)$, along with the appearance of the small-$k$ peak at $k=k_\lambda  \approx 1/\sigma$ which exhibits, in contrast, a slower but far more spectacular enhancement. 

The main peak around $k_{main}$ describes short-ranged correlations, of the order of one HS diameter, and implies rather fast processes of nearest neighbors moving close to each other in the initial stage of cluster formation. The small-$k$ peak at $k_\lambda$ describes, instead, longer-ranged correlations, of the order of six HS diameters, and its slower kinetics is associated with the build-up of cluster-cluster correlations, ultimately leading to dynamic arrest. This striking kinetic difference is visualized more precisely in the inset of Fig. \ref{Fig7a}, which compares $S_{main}(t;\phi,T)\equiv S(k_{main},t;\phi,T)$ with $S_\lambda(t;\phi,T)\equiv S(k_\lambda,t;\phi,T)$ for the 
same $t$ sequence. 

Let us stress again that the information provided by the full $t$-dependent solutions of the NE-SCGLE equations (\ref{dsktenst})-(\ref{fluctsquench}) is too abundant to be reviewed in one individual report, even if we restricted ourselves only to one specific property, such as to $S(k,t;\phi,T)$. The main reason is that this function depends on its four arguments, $(k,t;\phi,T)$, and it is not a simple task to scan this four-dimensional parameter space. Still, partial views of this dependence are particularly instructive. For example, 
Fig. \ref{fig6}(a) plotted  $S(k,t;\phi,T)$ as a function of $T$, keeping the other three arguments fixed, namely, $t=\infty$, $k=k_\lambda$, and $\phi=0.15$. Since we now are analyzing the dependence of $S(k,t;\phi,T)$ on the waiting time $t$, we may now extend such analysis to finite $t$.  This is done in Fig. \ref{fig7}(b), which plots $S_\lambda(t;\phi,T)$ for fixed $\phi=0.15$, as a function of $T$ for a sequence of values of the waiting time $t$. 

The main message of Fig.  \ref{fig7}(b) is that the NE-SCGLE theory not only predicts the results for $S_{\lambda}(t=\infty;\phi,T)=S^a_{\lambda}(\phi,T)$, which one would measure in an idealized and impossible experiment. A far more  relevant practical prediction refers to what one should measure at the \emph{finite} waiting times involved in real specific experiments or simulations. For instance, Fig. \ref{fig7}(b) illustrates the transient buildup of the structural correlations at $k_\lambda$, from which we learn that, although this is a slow process compared with the buildup of the correlations at $k_{main}$, it only takes a finite time, $t\approx 10^{3}$, for $S(k_\lambda,t;\phi,T)$ to saturate to its asymptotic value $S^a_{\lambda}(\phi,T)$ within the resolution of the figure. Hence, the temptation to identify this process with one of equilibration is enormous. Fortunately, the corresponding analysis of dynamic properties, such as the $\alpha$-relaxation time (not discussed in this paper) allows us to clearly discriminate between equilibration and arrest processes. Let us also mention a similarly interesting transient process, whose description can be drawn from  Fig. \ref{fig7}(a). We refer to the fact that $k_\lambda$ becomes
increasingly smaller with $t$, thus describing the (time-dependent)  cluster growth process.


The same structural information that we have just discussed can also be cast in terms of the NERDF $g(r,t;\phi,T)$. Here we shall not make any effort to review the resulting scenario, but only discuss  Fig.  \ref{fig7}(c), which is the real space counterpart of  Fig.  \ref{fig7}(a).  This figure describes the aging of $g(r,t;\phi,T)$ (along $\phi=0.15$), whose most relevant feature is the  slow emergence of the long-ranged correlations associated with cluster-cluster correlations, the main mechanism for dynamic arrest in the shallow quench considered. From the aging of $g(r,t;\phi,T)$, the non-equilibrium evolution of other relevant structural quantities (such as coordination numbers, effective cluster size, and volume fraction) could be derived, each of which adds a distinct perspective to the general scenario of cluster formation and dynamic arrest. As said above, however, in our present contribution we only aimed to pave the way to a more systematic discussion of both, structural  and dynamical properties. Such discussion is left for future work.

\section{Discussion and conclusions.}\label{section6}

In summary, in this paper, we have carried out the first systematic application of the non-equilibrium self-consistent generalized Langevin equation (NE-SCGLE) 
theory to the description of dynamic arrest in liquids with 
competing short-ranged attractions and long-ranged repulsions (SALR). For 
this, we have analyzed the non-equilibrium structural behavior in a model hard-sphere plus 
double Yukawa (HSDY) fluid, quenched into its region of  thermodynamical 
instability. 


For clarity, we have restricted ourselves to the study of the non-equilibrium structural evolution of the model type III SALR after a quench into the regions of thermodynamical instability.  The most fundamental quantity to understand such evolution is the thermodynamic stability function $\mathcal{E}(k;n,T)$. When interpreted as a familiar thermodynamic state function, it allows us to determine stability conditions, unstable domains and equilibrium structural properties. Its appearance in the kinetic equation (2), however, implies that $\mathcal{E}(k;n,T)$ bears a much deeper significance in determining more general non-equilibrium properties, such as the time-evolving structure factor $S(k;t)$. In this context, $\mathcal{E}(k;n,T)$ no longer represents only an equilibrium property. Instead, it becomes a fundamental input that determines the kinetics of the time-dependent structural correlations $S(k;t)$ and $g(r;t)$ (and of any other related property, such as those describing the non-equilibrium dynamics). In this sense, Eq. (2) can be viewed as an innovative proposal to extend the Ornstein Zernike equation to non-equilibrium conditions.

Restricted to the very specific set of parameters chosen in this study to represent the  type III SALR interactions ($z_1=1$, $z_2=0.5$ and $A=0.5$), the physical scenario predicted by the NE-SCGLE equations reveals a stunning and complex interplay between the thermodynamic instability 
represented by the so called $\lambda$ line $T=T_{\lambda}(\phi)$, and  
different underlying mechanisms for dynamical arrest. This interplay 
materializes in three distinct types of non-ergodic discontinuous (``type 
B") transitions: \emph{(i)} a ``\emph{fluid to arrested-cluster}" (F-AC) transition, 
occurring for concentrations below a threshold value $\phi_b=0.365$ and  
moderately low temperatures, which in practice coincides with the $\lambda$ line $T=T_{\lambda}(\phi)$; \emph{(ii)} an ``\emph{arrested-clusters to glass}" 
(AC-G) transition $T=T_c(\phi)$, occurring in the same $\phi$-regime but at lower 
temperatures, and; \emph{(iii)} a \emph{fluid to glass} (F-G) hard-sphere--like transition, observed for $\phi>\phi_b$, also denoted as  $T=T_c(\phi)$, to emphasize its smooth continuation inside the region of thermodynamic instability, as the previous AC-G transition (see Fig. \ref{Fig2}(b)). 

This asymptotic dynamic arrest transition scenario was greatly enriched by the discussion of the non-equilibrium structural properties represented by $S(k,t;\phi,T)$ and $g(r,t;\phi,T)$. We first discussed the asymptotic stationary limits $S^a(k;\phi,T)$ and $g^a(r;\phi,T)$, and then  considered  the non-equilibrium transient, provided by the full $t$-dependent solutions of the NE-SCGLE equations, Eqs. (\ref{dsktenst})-(\ref{fluctsquench}). Besides describing in some detail the most salient features, we also highlighted some seemingly innocuous observations. The most relevant of them refers to the fact that at first glance, and  for the specific HSDY model liquid considered in this work, the locus of the F-AC transition coincides \emph{``in practice''} with the $\lambda$ line $T=T_{\lambda}(\phi)$. These quotation marks needed a clarification, which was provided  by the zoom in the inset of Fig. \ref{fig6}(a): a closer inspection with the lens of the NE-SCGLE theory revealed that, in fact, the F-AC transition occurs not exactly at the  $\lambda$ temperature, but also at a dynamic arrest temperature $T_{(F-AC)}$ slightly higher, $T_{(F-AC)}(\phi)>T_{\lambda}(\phi)$. This immediately implied that  the F-AC transition preempts  the $\lambda$-instability, thus avoiding the divergence of the equilibrium structure factor expected to occur at the $\lambda$ line. As a consequence, when the model SALR system is quenched right below $T_{\lambda}$, the dynamical arrest conditions may prevent 
it from reaching full equilibrium conditions. 

This observation suggests the impossibility to experimentally observe the ordered inhomogeneous (``modulated") equilibrium phases predicted for the type III SALR systems in the region of thermodynamic instability. As just explained, the development of these phases is expected to become interrupted by non-equilibrium dynamical arrest barriers. Unfortunately, to the best of our knowledge no experimental results exist that can confirm (or disregard) these important predictions. Thus, our work could be a guide for future experiments on type III SALR systems, carried out to test the predicted non-equilibrium scenario just outlined. Similarly, the existing experimental data for type I and II SALR liquids surely will serve as a motivation for further characterizations within the theoretical framework provided by the NE-SCGLE. We leave such tasks for further work.

The present work was not aimed at scanning the parameter space  ($z_1$, $z_2$, $A$)  of the HSDY potential, that is, to explore the physical scenario for type I and type II SALR systems. Instead, it was meant  to pave the way to develop further systematic investigations with that and other purposes. An important pending issue refers to the fact that, within the NE-SCGLE description of arrested states, the specific evolution of the structural correlations is intimately related, via Eq.(2), to the  aging of the dynamics, with the former being represented in (2) by the mobility function $b(t)$. Since the specific kinetics of $S(k;t)$ determines the non-equilibrium relaxation of the dynamics, it is a pending task to characterize the aging of the dynamics along each of the transitions outlined in this work, and to establish its connection to the non-equilibrium structural behavior discussed here. From previous work on attractive systems (i.e., in the absence of the long-ranged Yukawa repulsion) we anticipate an intricate dynamic landscape, originated in  the complex interplay between thermodynamic instabilities and dynamical arrest, leading to distinct relaxation laws for the dynamics approaching each transition. All these aspects, however, shall be addressed in future work.

ACKNOWLEDGMENTS: This work was supported by the Consejo Nacional de Ciencia y Tecnología (CONACYT, México) through Postdoctoral Fellowship Grant No. I1200/224/2021 (L.F.E.-A.) and through Grants Nos. 320983, CB A1-S-22362, and LANIMFE 314881. The authors acknowledge the anonymous referees for their encouraging and constructive criticisms.

\pagebreak


\begin{thebibliography}{1}
\bibitem{shorttimed0}  The short-time self-diffusion coefficient  $D^0$ is related by Einstein's relation  $D^0=k_BT/\zeta^0$ with the  corresponding short-time friction coefficient $\zeta^0$ (determined by its Stokes expression \cite{landaulifshitzfm} in colloidal liquids, or by the kinetic [or `Doppler' \cite{ornsteinuhlenbeck}] friction coefficient in the case of molecular liquids  \cite{atomic1}).
\end{thebibliography}
\end{document}